\documentclass[a4paper, thmup to-restate]{lmcs}
\pdfoutput=1
\usepackage[utf8]{inputenc}

\usepackage{lastpage}
\lmcsdoi{20}{2}{11}
\lmcsheading{}{\pageref{LastPage}}{}{}%
{Nov.~07,~2022}{Jun.~05,~2024}{}

\usepackage{tikz}
\usepackage[linesnumbered,ruled,vlined]{algorithm2e}
\usepackage{colortbl,todonotes}

\author[S.~Akshay]{S. Akshay\lmcsorcid{https://orcid.org/0000-0002-2471-5997}}[a]
\author[H.~Bazille]{Hugo Bazille\lmcsorcid{https://orcid.org/0009-0000-4920-3819}}[b]
\author[B.~Genest]{Blaise Genest\lmcsorcid{https://orcid.org/0000-0002-5758-1876}}[c]
\author[M. Vahanwala]{Mihir Vahanwala\lmcsorcid{https://orcid.org/0009-0008-5709-899X}}[d]

\address{Indian Institute of Technology Bombay, Mumbai, India}
\email{akshayss@cse.iitb.ac.in}

\address{Laboratoire de Recherche de l'Epita (LRE), Rennes, France}
\email{hugo@lrde.epita.fr}

\address{CNRS, IRL 2955 IPAL, Singapore}
\email{blaise.genest@cnrs.fr}

\address{Max Planck Institute for Software Systems, Saarland Informatics Campus, Saarbr{\"u}cken, Germany}
\email{mvahanwa@mpi-sws.org}


\newcommand{\iu}{\mathrm{i}\mkern1mu}

\newcommand{\vect}[1]{\mathbf {#1}}

\newcommand{\zx}{\mathbb Z[X]}
\newcommand{\q}{\mathbb Q}
\newcommand{\size}{\text{size}}
\newcommand{\lrs}[1]{( #1 _n )_{n\in \mathbb{N}}}
\newcommand{\OO}{\mathcal{O}}
\newcommand{\Oone}{\mathcal{O}(1)}

\newcommand{\dominant}{\textnormal{dominant}}
\newcommand{\sOone}{s^{\Oone}}
\newcommand{\M}{\mathbf{M}}

\newcommand{\ptime}{\mathsf{PTIME}}
\newcommand{\pspace}{\mathsf{PSPACE}}
\newcommand{\np}{\mathsf{NP}}
\newcommand{\co}{\mathsf{co}}
\newcommand{\radius}{\psi}
\newcommand{\ball}{\mathcal{B}_\radius}
\newcommand{\cal}{\mathcal}

\makeatletter
\newenvironment{fakeabstract}{%
    \normalfont\Small
    \list{}{\labelwidth\z@
      \leftmargin3pc \rightmargin\leftmargin
      \listparindent\normalparindent \itemindent\z@
      \parsep\z@ \@plus\p@
      
    }%
    \item[] %
}{%
  \endlist
}
\makeatother

\title[ON ROBUSTNESS FOR SKOLEM, (ULTIMATE) POSITIVITY PROBLEMS]{On Robustness for the Skolem, Positivity\texorpdfstring{\\}{} and Ultimate Positivity Problems}
\titlecomment{Extended version of the conference paper that appeared in STACS'22.} \thanks{This work was partly supported by the DST/CEFIPRA/INRIA associated team EQuaVE, DST/SERB Matrices grant MTR/2018/00074, ANR-20-CE25-0012 MAVeriQ, and DFG grant 389792660 as part of TRR 248 (see https://perspicuous-computing.science).}

\begin{document}

\bibliographystyle{alphaurl}

\begin{abstract}
The Skolem problem is a long-standing open problem in linear dynamical systems: can a linear recurrence sequence (LRS) ever reach 0 from a given initial configuration? Similarly, the positivity problem asks whether the LRS stays positive from an initial configuration. Deciding Skolem (or positivity) has been open for half a century: the best known decidability results are for LRS with special properties (e.g., low order recurrences). On the other hand, these problems are much easier for ``uninitialised'' variants, where the initial configuration is not fixed but can vary arbitrarily: checking if there is an initial configuration from which the LRS stays positive can be decided by polynomial time algorithms (Tiwari in 2004, Braverman in 2006).

In this paper, we consider problems that lie between the initialised and uninitialised variants. More precisely, we ask if 0 (resp.\ negative numbers) can be avoided from every initial configuration in a neighbourhood of a given initial configuration. This can be considered as a robust variant of the Skolem (resp.\ positivity) problem. We show that these problems lie at the frontier of decidability: if the neighbourhood is given as part of the input, then robust Skolem and robust positivity are Diophantine hard, i.e., solving either would entail major breakthroughs in Diophantine approximations, as happens for (non-robust) positivity. Interestingly, this is the first Diophantine hardness result on a variant of the Skolem problem. On the other hand, if one asks whether such a neighbourhood exists, then the problems turn out to be decidable in their full generality, with $\pspace$ complexity. Our analysis is based on the set of initial configurations such that positivity holds, which leads to new insights into these difficult problems, and interesting geometrical interpretations. 
\end{abstract}

\maketitle

\begin{fakeabstract} 
\hspace*{\listparindent}Our techniques also allow us to tackle robustness for ultimate positivity, which asks whether there is a bound on the number of steps after which the LRS remains positive. There are two natural robust variants depending on whether we ask for a ``uniform'' bound on this number of steps, independent of the starting configuration in the neighbourhood. We show that for the uniform variant, results are similar to positivity. On the other hand, for the non-uniform variant, robust ultimate positivity has different properties when the neighbourhood is open and when it is closed. When it is open, the problem turns out to be tractable, even when the neighbourhood is given as part of the input.
\end{fakeabstract}

\section{Introduction}
A rational linear recurrence relation (LRR) of order $\kappa$ is
a relation $u_{n+\kappa} = \sum_{j=0}^{\kappa-1} a_j\cdot u_{n+j}$ defined by a tuple of coefficients $(a_0,\dots,a_{\kappa-1}) \in \mathbb{Q}^\kappa$, $a_0\neq 0$.  Given the initial configuration $(u_0,\ldots, u_{\kappa-1})$, i.e. the first $\kappa$ entries of the recurrence, which could be rationals or real algebraic numbers, there is a unique infinite sequence $\lrs u$ that satisfies the relation. This is called a Linear Recurrence Sequence (LRS).  The Skolem problem asks, given an LRS, i.e., a recurrence relation and an initial configuration, whether the sequence ever hits $0$,  i.e. does there exist $n \in \mathbb{N}$ with $u_n = 0$.  The positivity problem is a variant where the question asked is whether for all $n \in \mathbb{N}$, $u_n \geq 0$. Another variant is the ultimate positivity problem which asks whether there exists an integer $N_0\in \mathbb{N}$, such that for all $n\geq N_0$, $u_n\geq 0$. All these problems have applications in software verification, probabilistic model checking, discrete dynamic systems, theoretical biology, economics.

While the statements seem innocuous, the decidability of all these problems remains open since their introduction in the 1930's. Only partial decidability results are known, e.g., when the dimension is  less than 5 \cite{vereshchagin}. For a subclass of the so-called {\em simple} LRS, positivity is known to be decidable for order up to 9 \cite{ouaknine2014positivity}, while ultimate positivity is decidable for all orders for that class \cite{ouaknine2014ultimate}. On the other hand, the authors of \cite{joeljames3} prove an important hardness result: solving positivity or ultimate positivity would entail major breakthroughs in  Diophantine approximations. More precisely, one would be able to approximate the (Lagrange) type of many transcendental numbers, which deals with how close one can approximate the transcendental number using rational numbers having small denominators. 

This hardness result contrasts with positive results obtained for relaxations of the problems: instead of considering a fixed initial configuration, ~\cite{Tiwari04,Braverman06} consider every possible configuration as initial, i.e., they ask if there exists an initial configuration starting from which ensures that all entries of the sequence remain positive (this is sometimes called the uninitialised positivity problem). Surprisingly they show that this problem can be decided in $\ptime$. More recently, this result has been extended to processes with choices \cite{AGV18}. 

In this article, we consider natural variants that lie between the hard question of fixed initial configuration \cite{joeljames3}, and the easy question when the initial configuration is completely unconstrained \cite{Tiwari04,Braverman06}. Our goal is to undertake a comprehensive study of what happens when starting from a neighbourhood (ball) around the initial configuration. An immediate question that arises is whether the neighbourhood is part of the input or not and it turns out that this has a significant impact on decidability. Hence, we consider two sub-variants, first by fixing the neighbourhood and the second by asking if there exists a neighbourhood around the initial configuration (the existential variant). In both these cases, starting from any initial configuration in this neighbourhood we ask if

  \begin{itemize}
    \item all entries of the recurrence sequence remain positive. We call this the \emph{robust positivity problem}.
    \item  all entries of the recurrence sequence remain away from zero. We call this the \emph{robust Skolem problem}.
    \item all the entries of the recurrence sequence eventually (i.e., after a certain number of steps) become and remain positive. We call this the \emph{robust ultimate positivity problem}. In this last case, it also natural to consider a \emph{uniform variant}, namely whether there is a uniform bound $N_0$ on the number of steps, such that all starting configuration within the neighbourhood are positive after $N_0$.
  \end{itemize}

Our motivation to look at these problems stems from their role in capturing a powerful and natural notion of robustness, where the exact initial configuration cannot be fixed with arbitrarily high precision (which is often the case with real systems). 

We start by observing that as we need to tackle multiple initial configurations, we reason about the set of initial configurations from which positivity holds, which is sufficient to answer robustness questions. For this, we revisit the usual algebraic equations in a more graphical manner, which forms the crux of our approach. This allows us to reinterpret and generalise the hardness result of \cite{ouaknine2014positivity}, giving our first main contribution: if the neighbourhood is given as a fixed ball with real algebraic centre, then both robust Skolem and robust positivity are Diophantine hard, while robust ultimate positivity is Lagrange hard (these notions are defined formally in Section~\ref{sec:prelim}) in all cases except in the non-uniform case when the given ball is open. In this last case, it turns out that the problem can be solved in $\pspace$. Note in particular that the Diophantine hardness of robust Skolem is somewhat surprising, since it has recently been shown that the Skolem problem itself, at least in the case of simple LRS, can be solved assuming the Skolem conjecture and the p-adic Schanuel conjecture~\cite{BLNOPW22}, and hence is perhaps not expected to be Diophantine hard.

We then turn to the problems where the ball is not fixed, and ask if there exists a radius $\psi>0$ such that 0 or negative numbers can be avoided from every initial configuration in the $\psi$ ball around a given initial configuration. Our main contribution here is to show that the robust variants of the Skolem, positivity and ultimate positivity problems are all decidable in full generality, with $\pspace$ complexities. We summarise our results in Table~\ref{tab:results}, with the precise statements in Theorems~\ref{th.epsilonrobust},~\ref{th.robust} in Section~\ref{sec:problems}.

\begin{table}[t!]
\hspace*{-6pt}
\scalebox{0.94}{
  \begin{tabular}{|l|c|ll|c|}
  \hline
  & Exact & & Robust & {\bf$\exists$ Robust} \\
  \hline
  Skolem & \cellcolor{gray!25}
  ? (NP hard\cite{blondel2002}) & \cellcolor{red!25}&\cellcolor{red!25}{\bf Diophantine hard} & \cellcolor{green!25}\bf PSPACE*
  \\
  Positivity & \cellcolor{gray!25}Diophantine hard~\cite{joeljames3} & \cellcolor{red!25}&\cellcolor{red!25}Diophantine hard & \cellcolor{green!25}\bf PSPACE* 
  \\
 Ult. Pos &
  \cellcolor{gray!25}Lagrange hard~\cite{joeljames3}
  & Unif &\cellcolor{red!25}Lagrange hard & \cellcolor{green!25}\bf PSPACE\\
  &\cellcolor{gray!25}& Non-unif & \cellcolor{red!25}Lagrange hard (closed balls) & \cellcolor{green!25}\bf PSPACE\\
  &\cellcolor{gray!25}& & \cellcolor{green!25} {\bf PSPACE (open balls)} & \cellcolor{green!25}\bf PSPACE\\
  
  \hline
\end{tabular}
}
\caption{Complexity results for different variants of Skolem/Positivity/Uniform Positivity.  The most surprising results of our paper are in {\bf bold}. $\exists$ robustness is tractable in all cases. On the other hand, Robust Skolem is Diophantine hard while it is not the case for usual Skolem. Further, non-uniform robust ultimate positivity is tractable for open balls while every other variant of robust (ultimate) positivity shows some form of number theoretic hardness.
The asterisks (*) denote that these results hold for LRS of \textit{a priori} bounded order.}
  \label{tab:results}
\end{table}

\paragraph*{Related work} As mentioned earlier, the Skolem problem and its variants have received a lot of attention.  Given the hardness of these problems, $\varepsilon$-approximate solutions have been considered, e.g., in~\cite{BHM14,AAGT15} with different definitions of approximations. In comparison with our work, these are designed towards allowing approximate model checking. More recently, the notion of imprecision in Skolem and related problems was considered in~\cite{rounding20,pseudo21}. In \cite{rounding20}, the authors consider rounding functions at every step of the trajectory.  In \cite{pseudo21}, the so called Pseudo-Skolem problem is defined, where imprecisions up to $\varepsilon$ are allowed at every step of the trajectory, which is shown to be decidable in $\ptime$. These are quite different from our notion of robustness, which 
faithfully considers the trajectories generated from a ball 
representing $\varepsilon$-perturbations around the initial configuration. In a very recent extension~\cite{pseudo22} of~\cite{pseudo21}, it is shown that the existential robust Skolem question and a special case of the Pseudo-Skolem problem can be solved using o-minimality of the theory of reals with exponentiation. Lastly, \cite{N21} considers the problem when specified in a model of computation that takes real numbers, as opposed to rational numbers, as input. In this setting, perturbations are permitted in both the initialisation and the recurrence itself. This approach sidesteps several of the number-theoretic challenges, and decides problems on Linear Recurrence Sequences on all inputs except a set of measure 0.

  The novelty of this paper is three-fold: first, we provide the first comprehensive study of robustness focussed with respect to the initial configuration, covering several possible cases and variants; second, we provide and critically use geometric insight combined with the underlying number theory to prove our results; third, we show hardness results, in particular Diophantine hardness for a variant of the Skolem problem. This paper is an extended version of the work \cite{originalstacspaper} that featured in the proceedings of STACS 2022. The additional content can be summarised as follows:
\begin{enumerate}
\item We formalise the distinction between the notions of uniform and non-uniform Robust Ultimate Positivity.
\item We prove decidability results for non-uniform Robust Ultimate Positivity.
\item We leverage the underlying geometry to derive equivalences between robust Positivity, Uniform Ultimate Positivity and Skolem, in the cases of open and closed balls.
\item We prove Diophantine hardness of the robust problems at higher orders.
\end{enumerate}

\paragraph*{Structure of the paper.} The structure of this paper is as follows: In Section~\ref{sec:prelim} we define preliminaries, in particular the Skolem and (ultimate) positivity problem as well as known number-theoretic hardness results. In Section~\ref{sec:problems} we define the problem of our interest, namely robustness with respect to the initial configuration of Skolem as well as (ultimate) positivity and state all our main results. In Section~\ref{sec:geom} we provide a geometric interpretation of the behaviour of linear recurrences, which allows us to better understand and characterise the number-theoretic hardness results. In Section~\ref{sec:hardness}, we build upon the geometric insights to prove the hardness results claimed for robustness. In Section~\ref{sec:dec}, we again use and generalise the geometric interpretation in Section~\ref{sec:geom} to prove our positive results stated in Section~\ref{sec:problems} both in terms of decidability and complexity upper bounds. Finally we end with a conclusion in Section~\ref{sec:conclusion}.

\section{Preliminaries}
\label{sec:prelim}
Let $\kappa$ be any non-negative integer. We let $\mathbb{Q},\mathbb{R}$ denote the set of rationals and reals, respectively. Further, $\mathbb{Q}^\kappa,\mathbb{R}^\kappa$ denote $\kappa$-dimensional vectors over rationals, reals, respectively.
Let $\mathbf{c},\mathbf{d}$ be two vectors of $\mathbb{R}^{\kappa}$
that can be seen as one dimensional matrices of $\mathbb{R}^{\kappa\times 1}$. 
The distance between $\mathbf{c}, \mathbf{d}$ is defined as $||\mathbf{c} - \mathbf{d}|| =  \sqrt{(\mathbf{c} - \mathbf{d})^T(\mathbf{c} - \mathbf{d})}$, the standard $\ell^2$-distance. In this paper, we will consider two ways of ``measuring'' vectors: the first is the standard $\ell^2$-norm $||\mathbf{c}||$ for Euclidean length. The second is 
$\size(\mathbf{c})$, denoting the size of its bit representation i.e., number of bits needed to write down $\mathbf{c}$ (for complexity). We use the same notation for scalar constants with $\size(a)$ denoting the number of bits to represent an algebraic/rational constant $a$.
An algebraic number $\alpha$ is a root of a polynomial $p$ with integer coefficients. It can be represented \cite{mignottecon} by a 4-tuple $(p,a,b,r) \in \mathbb{Z}[X]\times \mathbb{Q}^3$ as the only root of $p$ at distance $<r$ from $a+ib$ (also see the Appendix). We define $\size(\alpha)$ as the size of the bit representation of $(p,a,b,r)$.

Given $\vect{c}\in\mathbb{R}^\kappa$, an {\em open (resp. closed) ball}  with centre $\vect{c}$ and  radius $\psi>0$, denoted $\mathcal B$, refers to the set of all vectors $\vect{c'}\in \mathbb{R}^\kappa$ such that $||\vect{c'}- \vect c|| < \psi$ (resp. $||\vect{c'}- \vect c|| \leq \psi$). For convenience we sometimes just say ball to mean an open or closed ball.

\subsection{Linear Recurrence Sequences}

We start by defining linear recurrence relations and sequences.

\begin{defi}
A linear recurrence relation, {\em LRR} for short, $\lrs{u}$ of order $\kappa$ is specified by a tuple of coefficients $\mathbf{a} = (a_0, \dots, a_{\kappa-1})$ with $a_0\neq 0$.
Given an initial configuration 
$\mathbf{c} = (c_0, \dots, c_{\kappa-1})$, 
the LRR uniquely defines a linear recurrence sequence (LRS henceforth), which is the sequence $(u_n(\vect c))_{n \in \mathbb{N}}$, inductively defined as $u_j(\vect c) = c_j$ for $j \leq \kappa-1$, and
$$u_{n+\kappa}(\vect c) = \sum_{j=0}^{\kappa-1}a_j u_{n+j}(\vect c) \text{ for all } n \in \mathbb{N}.$$
The companion matrix associated with the LRR/LRS (it does not depend upon the initial configuration $\vect c$) is:
$$\M = \begin{bmatrix}
	0 & 1 & 0 & \dots & 0 \\
	0 & 0 & 1 & \dots & 0 \\
	\vdots & \vdots & \vdots & \ddots & \vdots \\
	0 & 0 & 0 & \dots & 1 \\
	a_0 & a_1 & a_2 & \dots & a_{\kappa-1} 
\end{bmatrix}.$$ 
The characteristic polynomial of the LRR/LRS is 
$X^\kappa - \sum_{j=0}^{\kappa-1}a_jX^j$. 
The LRS is said to be simple if every root of the characteristic polynomial has multiplicity one. The size $s$ of the LRS is the size of its bit representation and is given by
$s = \sum_{j=0}^{\kappa - 1} \left(\size(a_j) + \size(c_j)\right)$. 

\end{defi}

When the coefficients, i.e., entries of $\mathbf{a}$, of an LRR are rational, we call it a {\em rational LRR}. {\bf In this paper, we will mostly be concerned with rational LRR, but the initial configuration $\mathbb{c}$ may have rational or real algebraic entries.} Note that arithmetic with algebraic numbers can indeed be performed with perfect precision: see the Appendix for references and a brief explanation.

Notice that given an initial configuration $\vect c$, we have that $\M^n \vect c = (u_n(\vect c),\ldots, u_{n+\kappa-1}(\vect c))$. Reasoning in the $\kappa$ dimensions $(u_n,\ldots, u_{n+\kappa-1})$ is a very useful technique that we will use throughout the paper as it displays the LRR as a linear transformation $\M$.

The characteristic roots of an LRR/LRS are the 
roots of its characteristic polynomial, and also the eigenvalues of the companion matrix. Let $\gamma_1, \dots, \gamma_r \in \mathbb{C}$ be the characteristic roots of the LRR/LRS.
An eigenvalue $\gamma_i$ is called {\em dominant} if it has
maximal modulus $|\gamma_i| = \max_{j \leq r} |\gamma_j|$, and {\em residual} otherwise. When $\mathbf{a}$ has rational entries, for all $j \leq r$, $\gamma_j$ is algebraic 
and $\size(\gamma_j) = s^{\Oone}$. We denote by $m_j$ the multiplicity of $\gamma_j$. We have $\sum_{j=1}^r m_j=\kappa$.

\begin{prop}[Exponential polynomial solution \cite{Everest2003RecurrenceS}]
\label{exppolysoln}
Given an initial configuration $\vect c$, there exists a unique tuple of coefficients $(\alpha_{ij}(\vect c))_{i \leq r, j < m_r}$ such that for all $n$, 
$$u_n(\vect c) = \sum_{i=1}^r \left(\sum_{j=0}^{m_r-1}\alpha_{ij}(\vect c) n^j \right) \gamma_i^n.$$ 
\end{prop}

The coefficients $\alpha_{ij}(\vect c)$ can be solved for from the initial 
state $\vect c$  \cite{halava}. When $\vect c$ has algebraic entries, it is implicit in the solution that for all $i, j$, both $\alpha_{ij}$ and $\frac{1}{\alpha_{ij}}$ are algebraic with values and norms upper bounded by $2^{\sOone}$. A formal proof of this claim can be found in \cite[Lemmas 4, 5, 6]{akshay_et_al}.

If the LRS is simple, then by definition $m_i=1$ for all $i$, and 
$u_n = \sum_{i=1}^r \alpha_{i}(\vect c) \gamma_i^n$, with 
$\alpha_{i}(\vect c)$ linear in $\vect c$, ie 
$\alpha_{i}(\vect \lambda \vect c+\lambda' \vect c')= \lambda \alpha_i(\vect c) + \lambda' \alpha_i(\vect c')$.

\begin{exa}
\label{ex.lrs}
As an example, consider the Linear Recurrence Relation of order 6 with 
$\vect a=(-1,4,-8,10,-8,4)$, i.e.
$u_{n+6}= 4 u_{n+5} - 8 u_{n+4} +10 u_{n+3} -8 u_{n+2} +4 u_{n+1} - u_n$. The roots of the characteristic polynomial are $1,e^{i 2 \pi \theta}, e^{-i 2 \pi \theta}$, with
$\theta = \frac{1}{3}$, 
each with multiplicity 2, and all dominant (they have the same modulus 1). 
The exponential polynomial solution is of the form 
$u_n(\vect c) = z(\vect c) n + z'(\vect c) + (x(\vect c) n+x'(\vect c)) e^{i 2 \pi n \theta} +
(y(\vect c) n + y'(\vect c)) e^{-i 2 \pi n\theta}$.
As $u_n(\vect c)$ is real, we must have that 
$x(\vect c),y(\vect c)$ are conjugates, as well as $x'(\vect c),y'(\vect c)$, and thus:
{\small
$$u_n(\vect c) = z(\vect c) n + z'(\vect c) + 
2( Re(x(\vect c)) n+ Re(x'(\vect c))) \cos(2 \pi n\theta) + 
2 (Im(x(\vect c)) n+ Im(x'(\vect c))) \sin(2 \pi n\theta).$$
}
\end{exa}

\subsection{Skolem and (ultimate) positivity problems}

\begin{defi}
Let $\lrs u$ be a rational LRR and $\vect c \in \q^\kappa$. The Skolem problem is to determine if there exists $n \in \mathbb{N}$ such that $u_n(\vect c)=0$. The positivity (resp. strict positivity) problem is to determine if for all $n \in \mathbb{N}$, $u_n(\vect c) \geq 0$ (resp. $u_n(\vect c)>0$). The ultimate positivity (resp. ultimate strict positivity) problem is to determine if there exists $N\in\mathbb{N}$ such that for all $n\geq N$, $u_n(\vect c) \geq 0 $ (resp. $u_n(\vect c)>0$).
\end{defi}

In this work, we will be more interested in the complement problem of  Skolem: namely, whether $u_n(\vect c) \neq 0$ for all $n$. This is of course equivalent in terms of decidability, but this formulation is more meaningful in terms of robustness, where we want to robustly avoid $0$.

The famous Skolem-Mahler-Lech theorem states that when $\vect c$ is algebraic, the set $\{i \, \mid \, u_i(\vect c)=0\}$ is the union of a finite set $F$ and finitely many arithmetic progressions \cite{skolem,mahler,berstel}.
These arithmetic progressions can be computed but the hard part lies in deciding if the set $F$ is empty: although we know that there is $N$ such that for all $n > N$, 
$n \notin F$, we do not have an effective bound on this $N$ in general. The
Skolem problem has been shown to be decidable for LRS of order up to $4$ \cite{mignotte,vereshchagin} and is still open for LRS of higher order. Also, only an $\np$ hardness bound is known if the order is unrestricted \cite{blondel2002,ABV17}.

For simple LRS, positivity has been shown to be decidable  up to order $9$ \cite{ouaknine2014positivity}. In \cite{ouaknine2014ultimate}, it is proved that positivity for simple LRS is hard for $\co\exists\mathbb R$, the class of problems whose complements are solvable in the existential theory of the reals.  A last result, from~\cite{joeljames3}, shows the difficulty of positivity, linking it to Diophantine approximations: how close one can approximate a transcendental number with a rational number with small denominator. We will follow the reasoning from \cite{joeljames3}. We start with two definitions.
  \begin{itemize}
    \item The \emph{Diophantine approximation type} of a real number $x$ is defined as:
$$L(x)= \inf \left\{ c \in \mathbb{R} \mid \left|x - \frac{n}{m}\right| < \frac{c}{m^2}, ~\text{for some } n,m \in \mathbb{Z}\right\}.$$
    \item The \emph{Lagrange constant} of a real number $x$ is defined as:
      $$L_\infty(x)= \inf \left\{ c \in \mathbb{R} \mid \left|x - \frac{n}{m}\right| < \frac{c}{m^2}, \text{ for infinitely many } n,m \in \mathbb{Z}\right\}.$$
  \end{itemize}
  
As mentioned in \cite{joeljames3}, the  Diophantine approximation type and Lagrange constant of most  transcendental numbers are unknown. Let $\mathcal A=\{p+q i \in \mathbb{C} \mid p,q \in \mathbb{Q} \setminus \{0\}, p^2+q^2=1\}$, i.e., the set of points on the unit circle of $\mathbb{C}$ with rational real and imaginary parts, excluding $1,-1, i$ and $-i$. The set $\mathcal A$ consists of algebraic numbers of degree 2, none of which are roots of unity \cite{joeljames3}. In particular, writing $p+q i= 2^{i 2 \pi \theta}=(-1)^{2\theta}$, we have that $\theta \notin \mathbb{Q}$ \cite{joeljames3}. We denote:
$$\mathcal{T} = \left\{ \theta \in (- 1/2, 1/2] \mid e^{2 \pi i \theta} \in \mathcal{A}\right\}.$$

  As argued in~\cite{joeljames3}, the set $\mathcal{T}$ is dense in $(- \frac 1 2, \frac 1 2]$, and consists solely of transcendental numbers. We assume that $\theta \in \mathcal{T}$ is specified by $p = \cos 2\pi i \theta$. In general, we don't have a method to compute $L(\theta)$ or $L_\infty(\theta)$ for $\theta \in \mathcal{T}$, or approximate them with arbitrary precision.

\begin{defi}
 We say that a problem is $\mathcal{T}$-Diophantine hard (resp. $\mathcal{T}$-Lagrange hard) if 
 its decidability entails that given any $\theta \in \mathcal{T}$ and $\varepsilon>0$ as input, 
 one can compute a number $\ell$ such that $|\ell - L(\theta)| < \varepsilon$ (resp. $|\ell -L_\infty(\theta)| <  \varepsilon$).
\end{defi}

Remarkably, in \cite{joeljames3}, it is shown that (i) if one can solve the positivity problem in general, then one can also approximate $L(\theta)$ and (ii) if one can solve the ultimate positivity, then one can approximate $L_\infty(\theta)$. That is, 

\begin{thmC}[\cite{joeljames3}]
\label{th.dioph}
Positivity for LRS of order 6 or above is $\mathcal{T}$-Diophantine hard and ultimate positivity for LRS of order 6 and above is $\mathcal{T}$-Lagrange hard.
\end{thmC}

\section{Robust Skolem,  Positivity and Ultimate Positivity}
\label{sec:problems}
The Skolem and (Ultimate) Positivity problems, as defined in the previous section, consider a single initial configuration $\vect c$. In this article, we investigate the notion of robustness, that is, whether the property is true in a neighbourhood of $\vect c$, which is important for real systems, where setting $\vect c$ with an arbitrary precision is not possible. We will consider two variants.  The first one fixes the neighbourhood as a ball $\mathcal B$, while the second asks for the existence of a ball $\mathcal B$ centred around a given initial configuration $\vect c$, such that for every initial configuration in $\mathcal B$, the respective condition is satisfied.

\begin{defi}[Robustness for Skolem, Positivity, Ultimate Positivity]
\label{defn:problems}
Let $(u_n)_{n \in \mathbb{N}}$ be the rational linear recurrence relation specified by a rational coefficient vector $\vect a$, and an initial algebraic configuration $\vect c$. 
Consider an algebraic ball $\mathcal B$ (with algebraic entries for both the centre and the radius). We define the following problems:

\begin{itemize}
 \item The {\em robust Skolem} problem is to determine if for all $\vect{c'}\in \mathcal B$ and all $n \in \mathbb{N}$,
 we have $u_n(\vect{c'}) \neq 0$.
 
 \item The {\em robust positivity} problem is to determine if for all $\vect{c'}\in \mathcal B$ and all $n \in \mathbb{N}$,
 we have $u_n(\vect{c'}) \geq 0$.

 \item The {\em robust non-uniform ultimate positivity} problem is to determine if for all $\vect{c'}\in \mathcal B$,  there exists $N_{c'} \in \mathbb{N}$ such that for all $n > N_{c'}$, we have $u_n(\vect{c'}) \geq 0$.
 
 \item The {\em robust uniform ultimate positivity} problem is to determine if there exists $N \in \mathbb{N}$ such that  for all $\vect{c'}\in \mathcal B$ and all $n > N$, we have $u_n(\vect{c'}) \geq 0$.
\end{itemize}

The $\exists$-robust variants of each problem asks whether there exists 
a ball $\mathcal B'$ centred around $\vect c$ such that the above holds over $\mathcal B'$.
\end{defi}

Note that for all the variants of $\exists$-robustness, there exists an open ball of radius $\psi>0$ for which robust Skolem (resp.\ positivity, uniform ultimate positivity) holds iff there exists a closed ball of radius $\psi'>0$ (e.g. $\psi'=\frac \psi 2$) for which it holds. Further, if there exists a ball with real radius, then there exist balls with algebraic and rational radii with the same centre. Thus we do not need to consider open and closed balls separately, nor do we need to explicitly mention the domain of the radius. For the other, i.e., {\em non} existential, variants of robustness as defined above, the case of closed and open balls can be different, and can also depend on whether the radius is rational or just real algebraic.

Our main results investigate the decidability and complexity of these problems.

\begin{thm}
  \label{th.epsilonrobust}
  For rational linear recurrence relations and algebraic balls:
    \begin{enumerate}
    \item for open and closed balls,
      \begin{enumerate}
      \item the robust positivity problem is $\mathcal{T}$-Diophantine hard,
      \item the robust Skolem problem is $\mathcal{T}$-Diophantine hard,
      \item the robust uniform ultimate positivity problem is $\mathcal{T}$-Lagrange hard
      \end{enumerate}
    \item for closed balls, the robust non-uniform ultimate positivity is $\mathcal{T}$-Lagrange hard.
    \end{enumerate}%
These lower bounds hold even 
for rational linear recurrence relations restricted to order 6 and 
for balls with rational radius.
\end{thm}

We remark that our proof of these lower bounds does not hold for balls whose centres have rational entries.

Our theorem above implies that for uninitialised positivity, one really needs the initial configuration to take a value possibly anywhere in the space rather than in a fixed neighbourhood to obtain decidability via \cite{Tiwari04, Braverman06}.  We remark that Diophantine hardness is known for the non-robust variant of positivity~\cite{joeljames3}, but to the best of our knowledge, it was not known for any variant of the Skolem problem. In fact, in light of the latest results in~\cite{BLNOPW22}, Diophantine hardness for (exact) Skolem seems unlikely, unless either of Skolem conjecture or p-adic Schanuel conjecture is falsified. {\color{black}We note that the p-adic techniques used therein rely on the input being integral, or rational, or algebraic. Our Diophantine hardness result, on the other hand, has connections to the positivity problem, and is intrinsically related to the common underlying geometry. In a nutshell, the distinction is that the robust variants of the problem implicitly reason about a continuum of initialisations, including those with transcendental coordinates, for which standard results like the Skolem-Mahler-Lech Theorem do not hold.}

Surprisingly, we obtain decidability for every linear recurrence relation and every initial configuration when considering a given {\em open} ball for non-uniform robust ultimate positivity, or by relaxing the neighbourhood to be as small as desired for any of the variants. This constitutes our second main result:

\begin{thm}
\label{th.robust}
The following decidability results hold for rational linear recurrence relations:
\begin{enumerate}
\item $\exists$-robust Skolem, $\exists$-robust positivity and $\exists$-robust (non)-uniform ultimate positivity are decidable for a centre $\vect c$ with {\em algebraic} entries. Further:
\begin{enumerate}
\item Deciding $\exists$-robust (non)-uniform ultimate positivity can be done in $\pspace$. 
\item When the centre $\vect c$ has {\em rational} entries, for any $K\in \mathbb{N}$, deciding $\exists$-robust Skolem and $\exists$-robust positivity for LRS of order at most $K$ can be done in $\pspace$.
\end{enumerate}
  \item Robust non-uniform ultimate positivity is decidable in $\pspace$ for open algebraic balls.
    \end{enumerate}
\end{thm}

The main difference between our techniques and several past works (except~\cite{STACS16} which is restricted to eigenvalues being roots of unity) is as follows: given an LRR $(u_n)_{n \in \mathbb{N}}$, our intuition and proofs hinge on representing the set $P$ of initial configurations $\vect d$ from which positivity holds. Formally:
$$P = \{\vect d \in \mathbb{R}^\kappa \mid u_n(\vect d) \geq 0 \text{ for all } n \in \mathbb{N}\}.$$

We may note that the set $P$ is convex. To see this, observe that for $d,d' \in P$,  for all $\alpha,\beta>0$ with $\alpha + \beta = 1$, we have $\alpha \vect d+\beta \vect d' \in P$  as $u_n(\alpha \vect d+\beta \vect d') = \alpha u_n(\vect d) + \beta u_n(\vect d') \geq 0$ for all $n$. We also remark that a definition similar to $P$ is possible for the set $S$ of initial configurations from which $0$ is avoided. But it turns out that that set is much harder to represent (e.g., it is not convex in general). Using $P$ surprisingly suffices to deal with robust Skolem as well.

In Section~\ref{sec:geom}, we provide the geometric intuitions behind our ideas as well as set up the notations for the proofs of the above theorems. We exploit the geometric intuitions from Section~\ref{sec:geom} in Section~\ref{sec:hardness}, to prove Theorem~\ref{th.epsilonrobust}. In Section~\ref{sec:dec} we prove Theorem~\ref{th.robust} providing algorithms for the decidable cases.

\section{Geometrical representation of an LRR for Diophantine hardness} 
\label{sec:geom}
{\color{black} The number-theoretic hardness of the non-robust variants starts at order 6; in this work, we show corresponding hardness results for the robust variants too. Decidability at lower orders is non-trivial: see \cite{vahanwala2023robust} for an exposition.} Hence, in this section and the next, we will focus on a particular LRR  of order $\kappa=6$, sufficient for the proofs of hardness, i.e. Theorem \ref{th.epsilonrobust}. In Section~\ref{sec:dec}, we will generalise some of the constructions explored here to obtain our decidability results stated in Theorem \ref{th.robust}.

Let $\theta \in \mathcal{T}$, i.e. $e^{i 2\pi \theta} = p+qi \in \mathcal A$, with both $p, q$ rational and $p^2+q^2=1$. 
We want to approximate $L(\theta)$ (indeed this is the problem that is ``Diophantine hard'').
For Lagrange hardness, we will adapt the construction and proof in section \ref{s.lagrange}, approximating $L_\infty(\theta)$ instead of $L(\theta)$.

Consider the Linear Recurrence Relation of order 6 defined by
$\vect a=(-1,4p+2,-(4p^2+8p+3),8p^2+8p+4,-(4p^2+8p+2),4p+2)$.
The roots of the characteristic polynomial are $1,e^{i 2 \pi \theta}, e^{-i 2 \pi \theta}$, each with multiplicity 2, and all dominant (they have the same modulus 1). Example 3 is a particular case of this $\vect a$, with $p=\frac{1}{2}= \cos(\frac{\pi}{3})$. However, notice that $\theta=\frac{1}{3} \notin \mathcal{T}$ as it corresponds to $q=\sin(\frac{\pi}{3})=\frac{\sqrt{3}}{2} \notin \mathbb{Q}$. Now, since $u_n(\vect c)$ is a real number for any $n$ and real initial configuration $\vect c$, we can write the exponential polynomial solution in the form:
{\small 
  \begin{align*}u_n(\vect c) = z_{dom}(\vect c) n - x_{dom}(\vect c) n \cos(2 \pi n\theta) - y_{dom}(\vect c) n \sin( 2 \pi n\theta)\\ + z_{res}(\vect c) - x_{res}(\vect c) \cos(2 \pi n\theta) - y_{res}(\vect c) \sin(2 \pi n\theta)
    \end{align*}
}
The coefficients $z_{dom}(\vect c),x_{dom}(\vect c),y_{dom}(\vect c)$ and $z_{res}(\vect c),x_{res}(\vect c),y_{res}(\vect c)$ are associated with the initial configuration $\vect c$ of the LRS. In the following, we reason in the basis of vectors $\overrightarrow{z_{dom}},\overrightarrow{x_{dom}},\overrightarrow{y_{dom}},\overrightarrow{z_{res}},\overrightarrow{x_{res}},\overrightarrow{y_{res}}$, as the geometrical interpretation is simpler in this basis. We will eventually get back to the original coordinate vector basis at the end of the process. 
From e.g., ~\cite[Section 2]{halava}, we know that we can transform from one basis to the other using an invertible Matrix $C$ with $C \cdot \vect c = (z_{dom}(\vect c),x_{dom}(\vect c),y_{dom}(\vect c),z_{res}(\vect c),x_{res}(\vect c),y_{res}(\vect c))$.

We study the positivity of $u_n$ by studying the positivity of $v_n=\frac{u_n}n$, for all $n \geq 1$. We denote 
$v_n^{dom}(z_{dom},x_{dom},y_{dom})= z_{dom} - x_{dom} \cos(2 \pi n\theta) - y_{dom} \sin(2 \pi n\theta)$,
which we call the dominant part of $v_n$, while we denote
$v^{res}_n(z_{res},x_{res},y_{res})=\frac{1}{n}(z_{res} - x_{res} \cos(2 \pi n\theta) - y_{res} \sin(2 \pi n\theta))$, which we call the residual part of $v_n$.
The residual part tends towards 0 when $n$ tends towards infinity because of the coefficient $\frac{1}{n}$.

\subsection{High-Level intuition and Geometrical Interpretation}
\label{sec:basicintuition}

We provide a geometrical interpretation of set $P$. We cannot characterise it exactly, even in this particular LRR of order $\kappa=6$ (else we could decide positivity for this case which is known to be Diophantine hard). To describe $P$, we define its ``section'' over $(z_{dom},x_{dom},y_{dom})$ given $(z_{res},x_{res},y_{res})$:
$$P_{(z_{res},x_{res},y_{res})} = \{(z_{dom},x_{dom},y_{dom}) \mid 
v_n(z_{dom},x_{dom},y_{dom},z_{res},x_{res},y_{res}) \geq 0 \text{ for all } n\}.$$

It suffices to characterise $P_{(z_{res},x_{res},y_{res})}$  for all $(z_{res},x_{res},y_{res})$ in order to characterise $P$, as
$P= \{(z_{dom},x_{dom},y_{dom},z_{res},x_{res},y_{res}) \mid (z_{dom},x_{dom},y_{dom}) \in P_{(z_{res},x_{res},y_{res})}\}$. Among these sets, one is particularly interesting: $P_{(0,0,0)}$, as it  is the set of tuples $(z_{dom},x_{dom},y_{dom})$ such that $v^{dom}_n(z_{dom},x_{dom},y_{dom}) \geq 0$ for all $n \in \mathbb{N}$. Our reason for focussing on this representation of $P$ is three-fold. First, unlike $P$, the set $P_{(0,0,0)}$ can be characterised exactly, as a cone depicted in Figure \ref{schema3} (this will be formally shown in Lemma \ref{cone} below). Second, the set $P_{(z_{res},x_{res},y_{res})}$ is in 3 dimensions that we can represent more intuitively than a 6 dimensional set. Last but not least, we can show that $P_{(z_{res},x_{res},y_{res})} \subseteq P_{(0,0,0)}$ for all $(z_{res},x_{res},y_{res})$ (Lemma \ref{Pres}).

\begin{figure}[!t]
\begin{center}
\begin{tikzpicture}

  
  \draw[] (-2,-0.5) -- (2,-0.5);
  \draw [] (-2,-0.5) -- (-1,0.5);
  \draw[] (-1,0.5) -- (3,0.5);
  \draw [] (2,-0.5) -- (3,0.5);

  
  \draw[] (0,0) -- (2.5,0);
  \draw[dashed] (2.5,0) -- (3.5,0);
  \draw[] (0,0) -- (1.5,2);
  \draw[dashed] (1.5,2) -- (2.1,2.8);
  \draw[rotate=-65.7] (-0.1,2.25)ellipse(1.12 and 0.3);
  
  
  \draw[->] (0,0) -- (3,1.5);
  \node (b0) at (3.25,1.25){$\overrightarrow{z_{dom}}$};
  
  \node (b1) at (0,-0.15){\tiny$(0,\ldots,0)$};

  \node (b2) at (1,-0.35){\tiny Hyperplane $H_1$};
\end{tikzpicture}
\end{center}

\caption{Visual representation of the cone $P_{(0,0,0)}$.}
\label{schema3}
\end{figure}
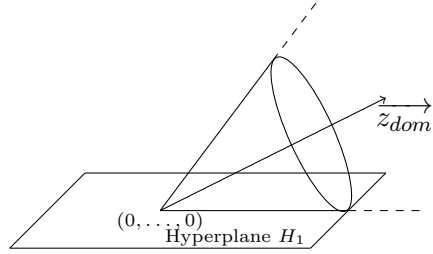

On the other hand, we also consider a related set in 6 dimensions: 
\begin{align} P_{dom} = \{(z_{dom},x_{dom},y_{dom},z_{res},x_{res},y_{res}) \mid \forall n, v^{dom}_n(z_{dom},x_{dom},y_{dom})) \geq 0\}.\label{eq:pdom}\end{align}
We note that $P_{(0,0,0)}$ is the projection of $P_{dom}$ over the 3 dimensions $(z_{dom},x_{dom},y_{dom})$. Also, characterising $P_{(0,0,0)}$ is sufficient to characterise $P_{dom}$ as 
$(z_{dom},x_{dom},y_{dom},z_{res},x_{res},y_{res}) \in P_{dom}$ iff 
$(z_{dom},x_{dom},y_{dom}) \in P_{(0,0,0)}$.
As $P_{(z_{res},x_{res},y_{res})} \subseteq P_{(0,0,0)}$ for all $(z_{res},x_{res},y_{res})$, we have $P \subseteq P_{dom}$.

We are now ready to represent $P_{(z_{res},x_{res},y_{res})}$
given some value $(z_{res},x_{res},y_{res})$.
We can interpret $P_{(z_{res},x_{res},y_{res})}$ in terms of half spaces:
$P_{(z_{res},x_{res},y_{res})}= \bigcap_{m=1}^{\infty} H^+_m(z_{res},x_{res},y_{res})$, with 
$H^+_m(z_{res},x_{res},y_{res})=\{(z_{dom},x_{dom},y_{dom}) \mid v_m(z_{dom},x_{dom},y_{dom},z_{res},x_{res},y_{res})) \geq 0 \}$.
The half space
$H^+_m(z_{res},x_{res},y_{res})$ is delimited by the hyperplane 
$$H_m(z_{res},x_{res},y_{res})=\{(z_{dom},x_{dom},y_{dom}) \mid v_m(z_{dom},x_{dom},y_{dom},z_{res},x_{res},y_{res})) = 0 \}.$$
which is a vector space ($\cos(2\pi m \theta)$ and $\sin(2\pi m \theta)$ are constant when $m$ is fixed). 

Consider the case of $(z_{res},x_{res},y_{res}) = (0,0,0)$.
We denote $H^+_m=H^+_m(0,0,0)$ and $H_m=H_m(0,0,0)$ for all $m$.
For instance, 
$H_0= \{(z_{dom},x_{dom},y_{dom}) \mid z_{dom}=x_{dom}\}$, as $v^{dom}_0(z_{dom},x_{dom},y_{dom})=z_{dom}-x_{dom}$. 

{\color{black}Define $\M_{dom}$ as the matrix that captures the action of the LRS 
$(v^{dom}_n)_{n \in \mathbb{N}}$ in the subspace of dominant coefficients of the exponential polynomial solution space.}
We have $H_{m}=\M_{dom} H_{m-1} = \M_{dom}^{m} H_0$.
We characterise $\M_{dom}$ in Lemma \ref{rotation} as a rotation 
around $\overrightarrow{z_{dom}}$ of angle $-2 \pi \theta$, which allows to characterise $H_m$ as the hyperplane which is the rotation of $H_0$ of angle $2 m \pi \theta$ around $\overrightarrow{z_{dom}}$.
That is, the cone shape for $P_{(0,0,0)}$is obtained by cutting away chunk of the 3D space delimited by hyperplanes $(H_m)$, the rotation $2n \pi \theta$ being dense in $[-\pi,\pi]$.

Coming back to some value $(z_{res},x_{res},y_{res}) \neq (0,0,0)$, we have that 
the hyperplane $H_n(z_{res},x_{res},y_{res})$ 
is parallel to the hyperplane $H_n$ (which is tangent to the cone $P_{(0,0,0)}$), because for $H_n$ of the form $u z_{dom} + v x_{dom} + w y_{dom} = 0$, we have $H_n(z_{res},x_{res},y_{res})$ is defined by 
$\{(z_{dom}, x_{dom}, y_{dom}) \mid $
$u z_{dom} + v x_{dom} + w y_{dom} = C\}$, for 
$C= \frac{z_{res} + x_{res} \cos(2 \pi n \theta) + y_{res} \sin(2 \pi n \theta)}{n}$ a constant as $n$ is fixed.

\begin{figure}[t!]
\centering
\includegraphics[scale=0.54]{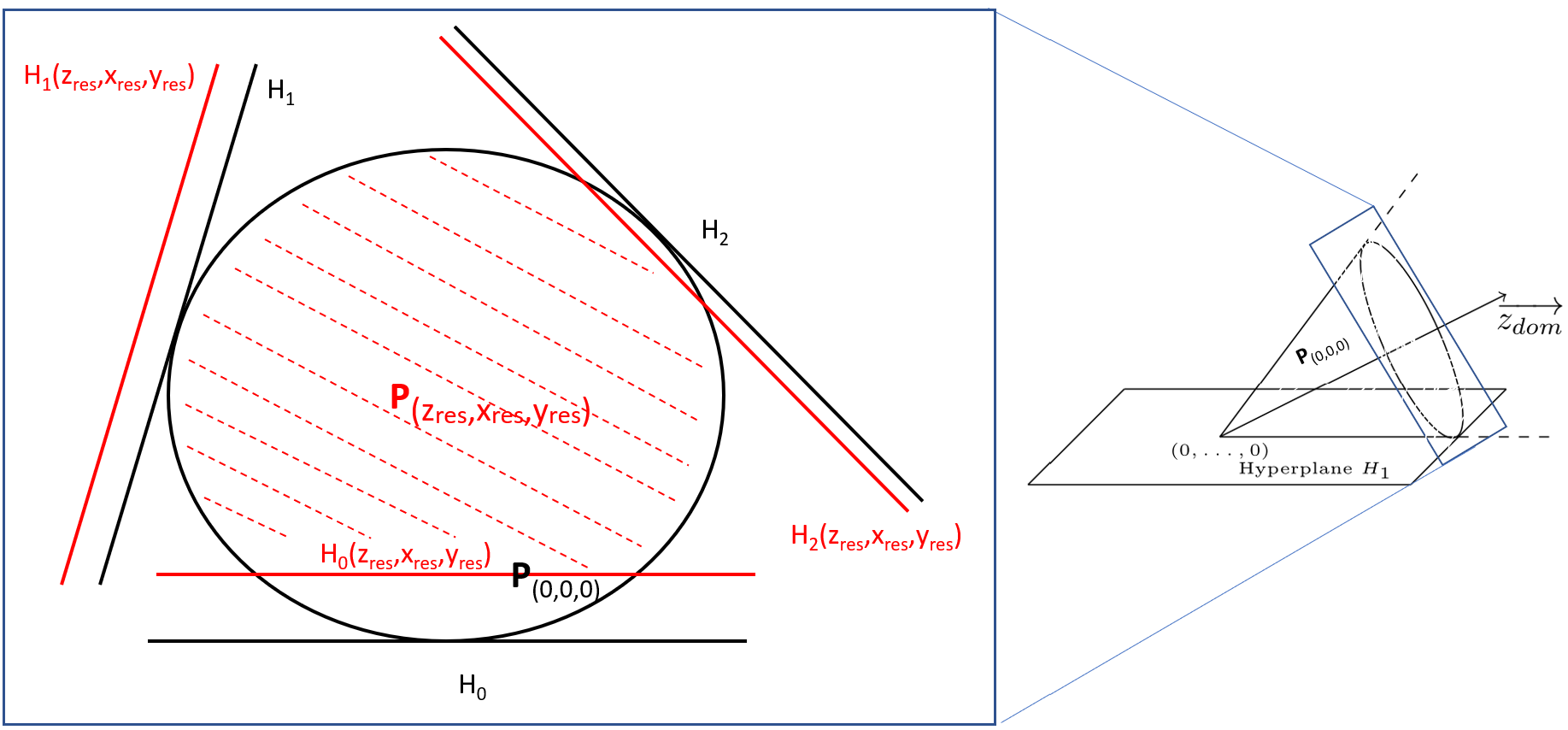}
\caption{Sections of 
$P_{(0,0,0)}$ (in black) and 
{\color{red} $P_{(z_{res},x_{res},y_{res})}$ (in dashed red)}, carved out  by hyperplanes $(H_i)$ (in black) and 
{\color{red} $(H_i(z_{res},x_{res},y_{res}))$ (in red)} respectively.}
\label{FPres}
\end{figure}

Thus, with this idea in mind, we can visualise $P_{(z_{res},x_{res},y_{res})}$ as depicted in Figure~\ref{FPres}, using $P_{(0,0,0)}$ and the hyperplanes $H_n(z_{res},x_{res},y_{res})$ parallel to $H_n$, with an explicit bound on the distance from $H_n(z_{res},x_{res},y_{res})$ to $H_n$, which further tends towards 0 as $n$ tends towards infinity. Next, we formalise the above intuition/picture into lemmas.

\subsection{Characterisation of $P_{(0,0,0)}$ and representing $P_{(z_{res},x_{res},y_{res})}$}
\label{sec:pdomcharac}
We now formalise some of the ideas in the above subsection. First, we start with Lemma \ref{cone} which shows that $P_{(0,0,0)}$ describes a cone, as displayed on Figure \ref{schema3}.

\begin{lem}
\label{cone}
$P_{(0,0,0)} = \{(z_{dom},x_{dom},y_{dom}) \mid z_{dom} \geq \sqrt{x_{dom}^2 + y_{dom}^2} \}$.
\end{lem}

\begin{proof}
We have $\cos(2 \pi n\theta)^2 + \sin(2 \pi n\theta)^2=1$ and 
$\cos(2 \pi n\theta)$ is dense in $[-1,1]$
as $\theta \notin \mathbb{Q}$. Denote $X=\cos(2 \pi n\theta)$, and study the function $f(X)= x_{dom} X + y_{dom} \sqrt{1-X^2}$.
Its derivative is $f'(X)= x_{dom} - \frac{y_{dom} X} {\sqrt{1-X}\sqrt{1+X}}$. 
We have $f'(X)=0$ iff $X=X_0=\frac{x_{dom}}{\sqrt{x^2_{dom} + y^2_{dom}}}$.
This gives us that $\max f(X) = f(X_0)=\frac{x^2_{dom} + y^2_{dom}}{\sqrt{x^2_{dom} + y^2_{dom}}}=\sqrt{x^2_{dom} + y^2_{dom}}$.
Thus, for all $(z_{dom},x_{dom},y_{dom})$ with 
$z_{dom} \geq \sqrt{x_{dom}^2 + y_{dom}^2}$, we have 
$z_{dom} \geq \max(f(X))$ and $v_n(z_{dom},x_{dom},y_{dom},z_{res},x_{res},$\\ $y_{res})\geq 
z_{dom} - f(X) \geq 0$ for all $n$.
On the other hand, if $z_{dom} < \sqrt{x_{dom}^2 + y_{dom}^2}$, then 
there exists $n$ such that $f(\cos(2 \pi n\theta))$ 
is arbitrarily close to $\max f(X) > z_{dom}$, and in particular 
$v_n = z_{dom} - f(\cos(2 \pi n\theta)) < 0$.
\end{proof}

We show now that the linear function $\M_{dom}$ associated with the LRR $(v^{dom}_n)_{n \in \mathbb{N}}$ 
is actually a rotation of angle $-2 \pi \theta$.

\begin{lem}
\label{rotation}
$\M_{dom}(z_{dom},x_{dom},y_{dom}) = 
(z_{dom},x_{dom} \cos(2\pi\theta)+ y_{dom} \sin(2\pi\theta), y_{dom} \cos(2\pi\theta) - x_{dom} \sin(2\pi\theta))$, that is 
$\M_{dom}$ is a rotation around axis $\overrightarrow{z}$ of angle 
$-2 \pi \theta$.
\end{lem}

\begin{proof}
  We use the formulas $\cos(a+b) = \cos(a)\cos(b) - \sin(a)\sin(b)$ and $\sin(a+b) = \sin(a)\cos(b) + \cos(a)\sin(b)$.
  
  Matrix $\M_{dom}$ transforms 
  $v^{dom}_n(z_{dom},x_{dom},y_{dom})$ into 
  $v^{dom}_{n+1}(z_{dom},x_{dom},y_{dom})$.
  Using the formulas above
  with $a=2\pi n \theta, b=2\pi \theta$, we have that for all $n \geq 1$, 
  $v^{dom}_{n+1}(z_{dom},x_{dom},y_{dom})=v^{dom}_{n}(z_{dom},x_{dom} \cos(2\pi\theta)+ y_{dom} \sin(2\pi\theta), y_{dom} \cos(2\pi\theta)-x_{dom} \sin(2\pi\theta))$
  for all $n$, and thus $\M_{dom}$ transforms $(z_{dom},x_{dom},y_{dom})$
  into $(z_{dom},x_{dom} \cos(2\pi\theta)+ y_{dom} \sin(2\pi\theta), y_{dom} \cos(2\pi\theta)-x_{dom} \sin(2\pi\theta))$.
  
  To see why the transformed coordinates are obtained by the rotation claimed above, observe the action in the plane of rotation. Indeed, consider a point $p$ in 2D space at Cartesian coordinates $(x_{dom},y_{dom})$.
  Its polar coordinates are $(r,\alpha)$, with $r=\sqrt{x_{dom}^2+y_{dom}^2}$ being the distance between $(0,0)$ and $p$.
  Consider the point at polar coordinates $(r,\alpha-2\pi \theta)$, i.e. the point obtained by rotating $p$ through an angle of $-2\pi\theta$.
  Through a straightforward application of trigonometric identities, we can see that it is at Cartesian coordinates $(r \cos(\alpha-2\pi \theta),r \sin(\alpha-2\pi \theta))= (r\cos(\alpha)\cos(2\pi \theta) + r\sin(\alpha)\sin(2\pi \theta) , 
  r\sin(\alpha)\cos(2\pi \theta) - r\cos(\alpha)\sin(2\pi \theta))
  =
  x_{dom} \cos(2\pi\theta) + y_{dom} \sin(2\pi\theta), y_{dom} \cos(2\pi\theta)- x_{dom} \sin(2\pi\theta))$.
On comparing with the form obtained in the preceding paragraph, we conclude that the rotation of angle $-2 \pi \theta$ transforms 
  $(x_{dom},y_{dom})$ into $(x_{dom} \cos(2\pi\theta)+ y_{dom} \sin(2\pi\theta), y_{dom} \cos(2\pi\theta)- x_{dom} \sin(2\pi\theta))$.
  \end{proof}

Finally, the following lemma implies that $P \subseteq P_{dom}$.

\begin{lem}
\label{Pres}
For all $z_{res},x_{res},y_{res}$, we have $P_{(z_{res},x_{res},y_{res})} \subseteq P_{(0,0,0)}$.
\end{lem}

\begin{proof}
We use the following simple but important observation. Let $(u_n)_{n\in \mathbb{N}}$ be an LRS where all roots have modulus 1, i.e., each root is of the form $\gamma= e^{\iu \theta}$, with distinct values of $\theta$. Let $u_j$ be the $j^{th}$ element of the LRS, with $j \in \mathbb{N}$.  Then for all $\varepsilon, N$, there exists $n>N$ with 
$|u_n-u_j| < \varepsilon$.
That is, for each value visited, the LRS will visit arbitrarily close values an infinite number of times. This is the case in particular of $v_n^{dom}$.

Now, assume for contradiction that there is a configuration $(z_{dom},x_{dom},y_{dom})$ in 
$P_{(z_{res},x_{res},y_{res})} \setminus P_{(0,0,0)}$.
Since $(z_{dom},x_{dom},y_{dom}) \notin P_{(0,0,0)}$, there exists $m$ with\\ $v^{dom}_m(z_{dom},x_{dom},y_{dom})<0$.
We let $\varepsilon = \frac{|v^{dom}_m(z_{dom},x_{dom},y_{dom})|}{3}$ and $N$ such that for all $n>N$, 
$|v^{res}_n|<\varepsilon$ (because it converges towards 0 when $n$ tends towards infinity).
From the above observation, we obtain an $n>N$ such that $|v^{dom}_n(z_{dom},x_{dom},y_{dom})-v^{dom}_m(z_{dom},x_{dom},y_{dom})| <\varepsilon$. Thus:
\begin{align*}
v_n(z_{dom},x_{dom},y_{dom},z_{res},x_{res},y_{res}) &=
v^{dom}_n(z_{dom},x_{dom},y_{dom}) + 
v^{res}_n(z_{res},x_{res},y_{res}) \\
&<v^{dom}_m(z_{dom},x_{dom},y_{dom}) + 
\varepsilon + \varepsilon \quad < 0.
\end{align*}
A contradiction with 
$(z_{res},x_{dom},y_{dom}) \in P_{(z_{res},x_{res},y_{res})}$.
\end{proof}

\section{Proof of Theorem \ref{th.epsilonrobust}}
\label{sec:hardness}

\subsection{Intuition for hardness of (robust) positivity}
\begin{figure}[t!]
\centering
\includegraphics[scale=0.54]{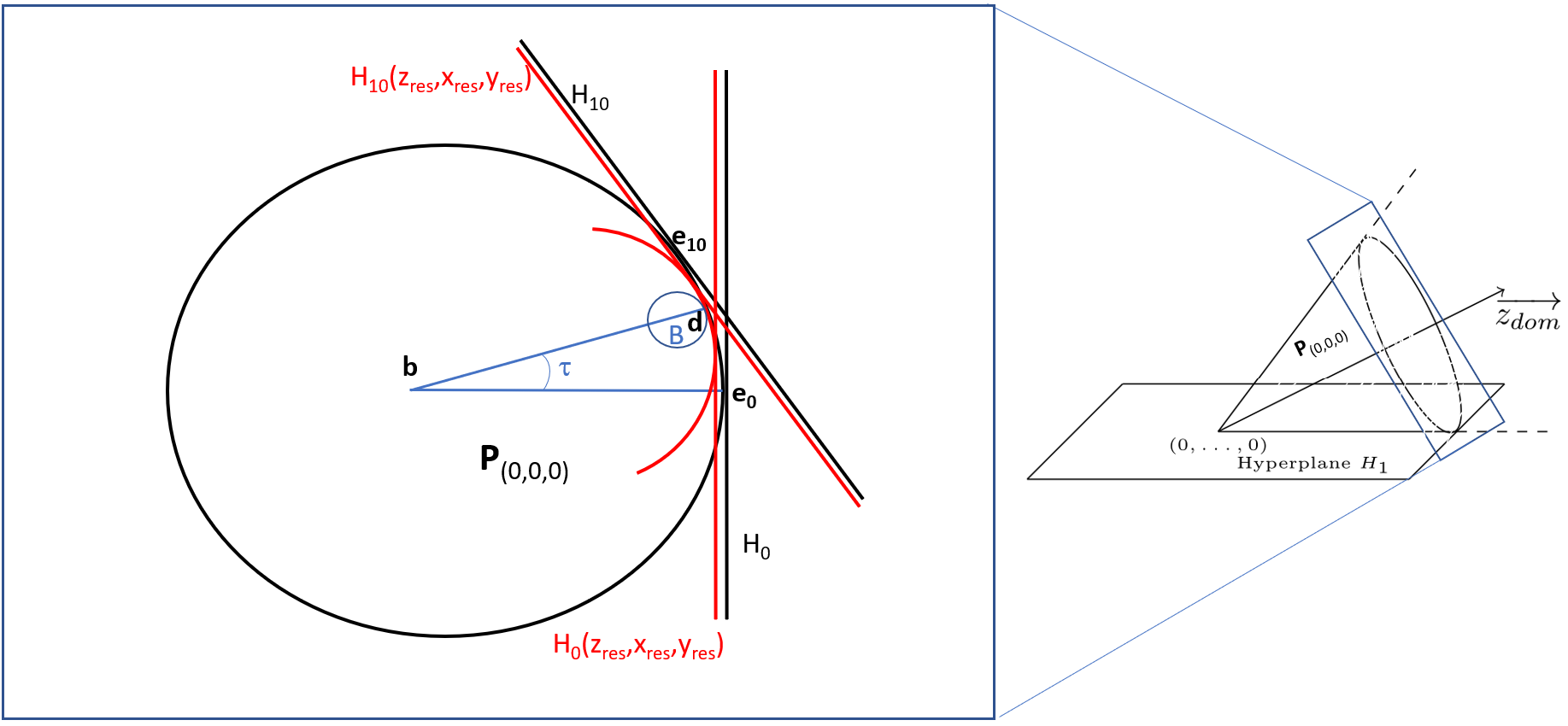}
\caption{Representation of a section of $P_{(0,0,0)}$, 
with hyperplanes $H_0,H_{10}$ being represented.}
\label{ballInt}
\end{figure}

Consider a vector $\vect d$ on the surface of $P_{dom}$, namely  $\vect d= (z_{dom},x_{dom},y_{dom},z_{res},x_{res},y_{res})$, that is, $(z_{dom},x_{dom},y_{dom}) \in P_{(0,0,0)}$. Consider the subset of $P_{(0,0,0)}$ which consists of points whose first coordinate $z_{dom}$ is the same as that of $\vect d$. For all $n$, let $\vect e_n$ be the point of this section where hyperplane $H_n$ is tangent to $P_{(0,0,0)}$.
Let $\tau$ be the angle made between the centre $b$ of the section, $\vect e_0$ and $\vect d$. Hence, $\vect e_0$ is at angle 0 and $\vect e_n$ at angle $2 \pi n \theta \mod 2 \pi$. We depict this pictorially in Figure~\ref{ballInt}.

We have that $u_n(\vect d) \geq 0$ for all $n$ iff  $\vect d$ is in the intersection of all half spaces defined by $H_i(z_{res},x_{res},y_{res})$.  
As $2 \pi n \theta \mod 2 \pi$ is dense in $[0,2\pi)$, 
for all $\beta>0$, there is a $n$ such that $\vect e_n$ is at angle $\alpha_n \in [\tau-\beta, \tau + \beta]$, hence
$H_n$ will be $\varepsilon$-close to $\vect d$. 
To know whether $\vect d$ is in the half space defined by $H_n(z_{res},x_{res},y_{res})$, we need to compare the distance $\varepsilon$ between $H_n$ and $\vect d$, with the value of $n$. If the value of $n$ is too large, then the distance between $H_n(z_{res},x_{res},y_{res})$ and $H_n$ is smaller than $\varepsilon$, and $\vect d$ is in the half space $H^+_n(z_{res},x_{res},y_{res})$.

In other words, for $(u_n(\vect d))_{n \in \mathbb{N}}$ not to be positive, $n$ needs to be both small enough and such that 
$2 \pi n \theta \mod 2 \pi$ is close to $\tau$. This is similar to $L(\theta)$ being small, as shown in Lemma \ref{l.positive}.

Now, for robust positivity (Theorem \ref{th.epsilonrobust}.1.a), we consider a ball $\mathcal B$ entirely in $P_{dom}$, tangent to the surface of $P_{dom}$ only on point $\vect d$. The ball will be positive iff the curvature of the ball is steeper than the curvature from hyperplanes  $H_n(z_{res},x_{res},y_{res})_{n \in \mathbb{N}}$ around $\vect d$, as shown in Lemma \ref{claim:fail}. This will correspond again to computing $L(\theta)$, thus showing hardness.

\subsection{Formalising the proof for closed balls and robust positivity}

In this section, we formalise the intuition given above, in the case of a closed ball and for robust positivity, i.e., Theorem~\ref{th.epsilonrobust}.1.a. We will extend this to the other cases of Theorem 
\ref{th.epsilonrobust} in the next subsection.

We recall the following:
 \begin{itemize}
    \item The \emph{Diophantine approximation type} of a real number $x$ is defined as:
$$L(x)= \inf \left\{ c \in \mathbb{R} \mid \left|x - \frac{n}{m}\right| < \frac{c}{m^2}, ~\text{for some } n,m \in \mathbb{Z}\right\}.$$
    \item The \emph{Lagrange constant} of a real number $x$ is defined as:
      $$L_\infty(x)= \inf \left\{ c \in \mathbb{R} \mid \left|x - \frac{n}{m}\right| < \frac{c}{m^2}, \text{ for infinitely many } n,m \in \mathbb{Z}\right\}.$$
  \end{itemize}

{\color{black}Now, for all $x \in \mathbb{R}$, we denote by $[x]$ the quantity $\min_{n \in \mathbb{Z}}|x - 2\pi n|$. Notice that $[x] \in [0,\pi]$; we follow the equivalent convention of \cite{dio-constants} to state that $L(\theta) = \frac{1}{2\pi}\inf_{n>0} n[2\pi n\theta]$, and that $L_\infty(\theta) = \frac{1}{2\pi}\liminf_{n>0} n[2\pi n\theta]$.

}

We show how to $\varepsilon$-approximate $L(\theta)$ in the following, using an oracle for robust positivity, following ideas in \cite{joeljames3}. To compute some $\ell$ that is $\varepsilon$-close to $L(\theta)$ for a given $\varepsilon >0$, we perform a binary search on $\ell$. 
{\color{black} By definition, it is clear that $L_\infty(\theta) \ge L(\theta) \geq 0$. An old observation of Dirichlet shows that every real number has Lagrange constant at most 1. Due to the work of Khintchine \cite{khintchine}, it is further known that these constants lie between $0$ and ${1}/{\sqrt{5}}$. So, for the binary search, we start with  a lower bound $\ell_{min}=0$ and an upper bound $\ell_{max}=1/\sqrt{5}$.}
For $\ell:=\frac {\ell_{min}+\ell_{max}} 2$, we want to know if $\ell \geq L(\theta) - \varepsilon$ (and then we set $\ell_{min}:=\ell$) or whether $\ell \leq L(\theta) + \varepsilon$ (and then we set $\ell_{max}:=\ell$).

\medskip

{\color{black}

Let  $\varepsilon >0$ and $\ell$ be a guess to check against $L(\theta)$. 
We define a {\em closed} ball $\mathcal B_{\psi}$ of radius $\sqrt{2} \psi$, centred at $\vect c=(2+\psi, 2-\psi, 0, 0, 0, 2 \pi \ell)$,  with $\psi<\frac 1 3$ and $\psi < {\pi \ell}$.
We now show that the ball $\mathcal B_{\psi}$ is entirely in $P_{dom}$,
only touching the surface of $P_{dom}$ at the point
$\vect d= (2, 2, 0, 0, 0, 2 \pi \ell)$:

\begin{lem}
  \label{l.ball}
 We have that
 $\vect d$ is on the surfaces of both $B_{\psi}$ and $P_{dom}$.
 
 Further, every $\vect d' =(z_{dom}, x_{dom}, y_{dom} , z_{res}, x_{res}, y_{res}) \in \mathcal B_{\psi} \setminus  \{\vect d\}$ satisfies 
 \begin{itemize}
  \item $x_{dom} < z_{dom}$, and
  \item $\vect d'$ is in $P_{dom}$ but not at the surface of $P_{dom}$.
 \end{itemize}
\end{lem}

\begin{proof}
First, $\vect d=(2, 2, 0, 0, 0, 2 \pi \ell)$ satisfies $||\vect c - \vect d|| = \sqrt{2} \psi$, so $\vect d \in B^{\psi}$, on the surface of $B_{\psi}$. 
Further, $\vect d$ satisfies the equation of Lemma \ref{cone}:
$2^2= 4 = (2+0)^2$, so $\vect d \in P_{dom}$ at its surface.

\medskip
Let $\vect d' =(z_{dom}, x_{dom}, y_{dom} , z_{res}, x_{res}, y_{res}) \in \mathcal B_{\psi} \setminus  \{\vect d\}$.

Thus, $(z_{dom} - (2+\psi))^2 + (x_{dom} - (2-\psi))^2 + y^2_{dom} 
+z^2_{res}+ x^2_{res}+ (y_{res}-2 \pi \ell)^2 \leq 2 \psi^2$.
In particular, we have $(z_{dom} - (2+\psi))^2 + (x_{dom} - (2-\psi))^2 +y_{dom}^2 \leq 2 \psi^2$, i.e. $(z_{dom}-2)^2 + (x_{dom}-2)^2 + y_{dom}^2 \leq 2 \psi (z_{dom} - x_{dom})$.
That is, $x_{dom} \leq z_{dom}$ as $(z_{dom}-2)^2 + (x_{dom}-2)^2 + y_{dom}^2  \geq 0$. Further, equality $z_{dom} = x_{dom}$ would imply that $z_{dom}=x_{dom}=2$ and $y_{dom}=0$> Since the ball is centred at $(2 +\psi, 2-\psi, 0, 0 , 0, 2\pi\ell)$ and has radius $\sqrt{2}\psi$, this implies that $\vect d'=\vect d$, which is not possible. Hence $x_{dom} < z_{dom}$.

\medskip

For the last result, by Lemma \ref{cone}, it suffices to prove that 
 $x_{dom}^2 + y^2_{dom} < z_{dom}^2$. Assume by contradiction that 
 $z_{dom}^2 - x_{dom}^2 - y^2_{dom} \leq 0$.
As $\vect d' \in \mathcal B_{\psi}$, we have in particular
$(z_{dom} - 2 - \psi)^2 + (x_{dom} - 2 + \psi))^2 + y^2_{dom} \leq 2 \psi^2$.
Adding the two inequalities, we obtain:
$$(z_{dom} - 2)^2 + z_{dom}^2  - 2 \psi (z_{dom} - 2) 
+ 2x_{dom}(\psi-2) + 4 - 4 \psi \leq 0.$$

\noindent That is $(z_{dom} - 2)^2 + (z_{dom} - 2)^2 + 4 z_{dom}
- 2 \psi z_{dom} + 2 x_{dom} (\psi-2)\leq 0$.

\noindent Hence $2(z_{dom} - 2)^2  + 2 (z_{dom} - x_{dom}) (2-\psi)\leq 0$,
a contradiction: indeed, the sum of a positive term and a strictly positive term (
$x_{dom} < z_{dom}$ by the above statement) cannot be negative or zero.
\end{proof}

In other words, $\vect d$ is the only point where the ball $\mathcal B_\psi$ intersects the surface of $P_{dom}$.
}

We now explain the relationship between the positivity of 
$(u_n(\vect d))_{n \in \mathbb{N}}$ and $L(\theta)$, which is the crux of the proof of Theorem \ref{th.dioph} by \cite{joeljames3}.

\begin{lem}
  \label{l.positive}
  For every $\ell, \varepsilon > 0$, we can compute an $n_1 >0$ such that for all $n\geq n_1$,
  we have $u_n(\vect d) \geq 0$ and 
  $2 \pi n \theta \mod{2\pi} \in [0,\pi]$
  together imply
  $n [2 \pi n \theta] > 2\pi\ell - \varepsilon$.
  \end{lem}

  \begin{proof}
    Assume that $u_n(\vect d) \geq 0$ and $2 \pi n \theta \mod{2\pi} \in [0,\pi]$.
    First, notice that as $2 \pi n \theta \mod{2\pi} \in [0,\pi]$, we have 
    $$u_n(\vect d) = 2n - 2 n \cos([2 \pi n \theta]) 
    - 2 \pi \ell \sin([2 \pi n \theta]).$$
    
    Let $\alpha_n= [2 \pi n \theta] \geq 0$. Considering the Taylor development 
    for $\alpha_n $ close to $0$ of $(1-\cos(\alpha_n))$
    and $\sin(\alpha_n)$, we get 
    $u_n(\vect d)/n= \frac{2}{2}\alpha_n^2 - \frac{2 \pi \ell \alpha_n}{n} + f(\alpha_n)$, with $f(\alpha_n)=O(\alpha_n^3)$.
    We have $u_n(\vect d) \geq 0$, thus
    $\frac{2 \pi \ell \alpha_n}{n}$ is smaller than 
    $\alpha_n^2 (1+\frac{f(\alpha_n)}{\alpha_n^2})$, that is 
       $\alpha_n \geq \frac{2 \pi \ell}{n(1+\frac{f(\alpha_n)}{\alpha_n^2})}$.
    There exists a computable value $\alpha_0 >0$ such that $\alpha_n<\alpha_0$ implies $1 - \frac{\varepsilon}{2\pi \ell} \leq  \frac{1}{(1+\frac{f(\alpha_n)}{\alpha_n^2})} \leq
    1 + \frac{\varepsilon}{2\pi\ell}$. 
    That is, if $\alpha_n \leq \alpha_0$, then 
    $n [2 \pi n \theta] > 2\pi\ell - \varepsilon$ and we are done.

  We define $n_1= \lfloor \frac{2\pi \ell - \varepsilon}{\alpha_0}\rfloor+1$, and thus
  $n_1 \alpha_0 > 2\pi \ell - \varepsilon$.
Hence if $n>n_1$ and $\alpha_n > \alpha_0$, we also have
$2\pi\ell - \varepsilon < n \alpha_n$.
\end{proof}
    
We reason about the case
$2 \pi n \theta \mod{2\pi} \in [\pi, 2\pi)$ symmetrically with respect to the case $2 \pi n \theta \mod{2\pi} \in [0, \pi)$:
it suffices to consider the configuration $\vect d^-=(2, 2, 0, 0, 0, -2 \pi \ell)$. 
If both $u_n(\vect d) \geq 0$ and $u_n(\vect d^-)\geq 0$, then $n [2\pi n \theta] > 2\pi \ell - \epsilon$.
The configuration $\vect d^-$ can be handled by considering the ball $\mathcal B^-_\psi$
of radius $\sqrt{2} \psi$ centred in $\vect c^- = (2+\psi, 2-\psi, 0, 0, 0, -2 \pi \ell)$.

\medskip

We turn now to the positivity of the balls $\ball \cup \ball^-$. 
We show that $\psi$ can be chosen small enough such that the following is true:

\begin{lem}
  \label{claim:fail} 
  For any $\ell, \varepsilon$, let $\vect d, \vect d'$ be defined as above. 
  Then for any $n_1$, we can choose $\psi$ small enough and $n_2 > n_1$ such that
  for all $n > n_2$, if $n [2\pi n \theta] \geq 2 \pi \ell + \varepsilon$, then $u_n(\vect d') \geq 0$ for all $\vect d' \in \ball \cup \ball^-$.
  \end{lem}
  
\begin{proof}
We first consider the initial configuration  $\vect d' \in  \ball \cup \ball^-$
which minimises $u_n(\vect d')$:

\begin{clm}
  $$\min_{\vect d' \in \ball \cup \ball^-} u_n(\vect d') =
  n(2-\radius)(1- \cos (2\pi n\theta)) - 2\pi\ell |\sin (2\pi n\theta) |
  -2\radius(\sqrt{n^2+1}-n). $$
\end{clm}
  
  \begin{proof}
    Any point $\vect d' \in \ball$ can be expressed as $\mathbf{c} + \mathbf{r}$, where $\mathbf{c}$ is the centre of one of the two spheres, and $\mathbf{r}$ is an arbitrary vector whose length does not exceed $\sqrt{2}\radius$. We know that $u_n(\vect d') = \mathbf{x_n}^T(\mathbf{c}+ \mathbf{r})$, where $\mathbf{x_n} = [n, n\cos (2\pi n\theta), n\sin (2\pi n\theta), 1, \cos (2\pi n \theta), \sin (2\pi n\theta)]^T$ is fixed. As discussed in the previous lemma, the minimum contribution from $\mathbf{x_n}^T\mathbf{c}$ is (note that the choice is over the centres of the two spheres)
    $$
    n(2+\radius) - n(2-\radius)\cos (2\pi n\theta) - 2\pi\ell |\sin (2\pi n\theta)|
    $$
    which can also be rearranged as
    $$
    2n\radius + n(2-\radius)(1-\cos (2\pi n\theta)) - 2\pi\ell |\sin (2\pi n\theta)|.
    $$
    It now remains to independently optimise over $\mathbf{r}$. For this, we note $\mathbf{x_n}^T\mathbf{r}$ is minimised when $\mathbf{r}$ has longest possible length, and is oriented opposite to $\mathbf{x_n}$. In this case, $\mathbf{x_n}^T\mathbf{r}$ will be the negative product of the lengths of $\mathbf{x_n}$ and $\mathbf{r}$. This is $-\sqrt{2n^2 + 2}\cdot\sqrt{2}\radius$, which simplifies to
    $$
    -2\radius\sqrt{n^2+1}.
    $$
    Adding the two contributions gives the result.
  \end{proof}

Thus, we have
$ \frac{u_n(\vect d')}{n} = (2 - \radius)(1 - \cos (2\pi n\theta)) - \frac{2\pi\ell}{n}|\sin (2\pi n\theta)| - 2{\radius}\left({\sqrt{1 + \frac{1}{n^2}} - 1}\right)$.

We let $\alpha_n=[2\pi n \theta] \ge 0$. Note that $|\sin \alpha_n| = \sin [\alpha_n]$

We first show that if $n \alpha_n  > 36 \pi$ then $\ball \cup \ball^-$ is positive, i.e. $\vect d'$ is:

\begin{clm}
  If $\alpha_n n > 36 \pi$, then $u_n(\vect d') > 0$ for all $\vect d' \in \ball \cup \ball^-$.
\end{clm}

\begin{proof}
Assume that $\alpha_n n > 36 \pi$.
We use $\psi \leq 1$, $\sqrt{1+y}-1 \leq \frac{y}{2}$,
$\sin y \leq y$ and $1-\cos y \geq \frac{y^2}{2} - \frac{y^4}{24}$ to obtain:
\begin{align}
  \label{eq1}
  \quad \, \quad \, \quad \, \quad \frac{u_n(\vect d')}{n} & \geq \frac{\alpha_n^2}{2} 
  - \frac{\alpha_n^4}{24} - \frac{2\pi\ell \alpha_n}{n}- \frac{\radius}{n^2}.
\end{align}

As $\alpha_n \leq \pi$, we have $\alpha_n^2 <10$ and 
$\frac{\alpha_n^2}{2}   - \frac{\alpha_n^4}{24} \geq \frac{\alpha_n^2}{12}$. That is:
\begin{align*}
  \quad \, \quad \, \quad \, \quad \frac{u_n(\vect d')}{n} & \geq \alpha_n (\frac{\alpha_n}{12} -  
  \frac{2\pi\ell}{n})- \frac{\radius}{n^2}.  
\end{align*}

As $\ell \leq 1$ and 
$\alpha_n \geq \frac{36 \pi}{n}$, we have: 
\begin{align*}
  \quad \, \quad \, \quad \, \quad \frac{u_n(\vect d')}{n} & \geq 
  \alpha_n (\frac{3 \pi}{n} - \frac{2\pi}{n})- \frac{\radius}{n^2} \\
  & \geq \frac{36 \pi^2}{n^2} - \frac{\radius}{n^2}.
\end{align*}

Hence, as $\psi \leq 1 < 36 \pi^2$, we have $u_n(\vect d')>0$.
\end{proof}

Hence, we can assume without loss of generality that 
$\alpha_nn \leq 36 \pi$.

\medskip

Assume now that $2\pi \ell + \varepsilon \leq \alpha_nn \leq 36 \pi$.
We want to show that $u_n(\vect d')\geq  0$.
Assume by contradiction that $\frac{u_n(\vect d')}{n} < 0$.
Similarly as (\ref{eq1}), we have:
\begin{align*} 
  \quad \, \quad \, \quad (2-\psi) (\frac{\alpha_n^2}{2} - \frac{\alpha_n^4}{24}) - \frac{2\pi\ell \alpha_n}{n}- \frac{\radius}{n^2} \leq  \frac{u_n(\vect d')}{n} < 0.
\end{align*}

We choose $\tau_1 = \frac{2 \pi \ell} {2 \pi \ell + \frac{\varepsilon}{3}} <1$ and $0 < \psi \leq 2 - 2 \tau_1$. We get:
$$
\tau_1 \alpha_n^2 - \tau_1 \frac{\alpha_n^4}{12} 
< \frac{2\pi\ell}{n} \alpha_n + \frac{\psi}{n^2}.
$$

We multiply by $\frac{n}{\tau_1 \alpha_n}>0$ and obtain:
$$
\alpha_n n  <
\frac{2\pi\ell}{\tau_1} + \frac{\alpha_n^3 n}{12} 
+ \frac{\psi}{\tau_1 \alpha_n n}.
$$

Now, $\frac{1}{\alpha_nn} \leq \frac{1}{2 \pi \ell}$,
$\alpha_n^3 n = O({\frac{1}{n^2}})$ as
$\alpha_n \leq \frac{36 \pi}{n}$, 
and choosing $n_2$ with $\left(\frac{36\pi}{n_2}\right)^3 n_2 \leq \frac{12 \epsilon}{3}$
and $\psi$ with $\psi < \frac{2 \tau_1 \pi \ell \epsilon}{3}$, we obtain 
the contradiction:
\begin{align*}
\alpha_nn < (2 \pi \ell + \frac{\varepsilon}{3}) + \frac{\varepsilon}{3} + \frac{\varepsilon}{3} = 2 \pi \ell + \varepsilon. \tag*{\qedhere}
\end{align*}
\end{proof}
 
 Now, assuming we have an oracle for $\psi$-robust positivity, we shall show how to use Lemma \ref{l.positive} (along with its symmetric analog for $\ball^-$) and Lemma \ref{claim:fail} to deduce either $L(\theta) < \ell+\varepsilon$ or $L(\theta) > \ell-\varepsilon$. These lemmata directly allow us to also reason about $\liminf n[2\pi n\theta]$ just the way we do about $\inf n[2\pi n\theta]$, i.e.\ the reduction from computing $L_\infty(\theta)$ to uniform $\psi$-robust ultimate positivity (Theorem~\ref{th.epsilonrobust}.1.c) is a straightforward adaptation of the proof we give of Theorem~\ref{th.epsilonrobust}.1.a.

\begin{proof}[Proof of Theorem \ref{th.epsilonrobust} for robust positivity and closed balls]
Let $\varepsilon >0$.
Assume that an $\ell$ has been fixed, such that we want to know either $L(\theta)< \ell+\varepsilon$ or $L(\theta)> \ell-\varepsilon$.
First, we fix $n_2$ and $\psi < \frac{1}{3}$ and $\psi < \pi \ell$ as per Lemma \ref{claim:fail}. The high level plan is to use the robust positivity oracle to query whether neighbourhoods that are subsets of $\mathcal B_\psi$ (symmetrically $\mathcal B_\psi^-$) containing $\vect d$ (symmetrically $\vect d^-$) are positive from iterate $n_2$ onwards. If yes, it means that $\vect d$ and $\vect d^-$ in particular are positive, and we use Lemma \ref{l.positive} and its symmetric statement to argue $L_{\ge n_2}(\theta) > \ell - \varepsilon$. If not, it means that the larger $\mathcal B_\psi$ and $\mathcal B_\psi^-$ themselves are not positive, and we use the contrapositive of Lemma \ref{claim:fail} to argue that $L_{\ge n_2}(\theta) < \ell + \varepsilon$. 

To complete the reduction, we need to decide whether $L_{\le n_2}(\theta) = \frac{1}{2\pi}\inf_{0 < n \le n_2} n[2\pi n\theta] < \ell + \varepsilon$. This is straightforward as it only 
involves computing $n[2\pi n \theta]$ to sufficient precision for a bounded number of indices $n$. 
If $L_{\leq n_2}(\theta) < \ell+\varepsilon$, 
then we know 
$L(\theta) \leq L_{\leq n_2}(\theta) < \ell+\varepsilon$.

We remark that $\mathcal B_{\psi}$ (symmetrically $\mathcal B_\psi^-$) corresponds to a ball in the coordinates\\ $(z_{dom},x_{dom},y_{dom},z_{res},x_{res},y_{res})$ of the coefficient space. In fact, as outlined in the plan above, our reduction needs to supply the initialisation $(v_{n_2}, \dots, v_{n_2+5})$ as input in the original coordinates $(v_0,v_1,v_2,v_3,v_4,v_5)$. The ball $\mathcal B_{\psi}$ is mapped to an hyper-ellipsoid $\mathcal O$ in the original coordinates. We can explicitly define a smaller ball $\mathcal B' \subseteq \mathcal O$ in the original coordinates, containing (the image of) $d \in \mathcal B'$.
Hence $\mathcal B$ is positive implies that $\mathcal B'$ is.

We symmetrically do the same thing for $\ball^-$, and query both neighbourhoods, thus implementing the reduction as outlined above.

\smallskip

{\color{black}
Finally, we show the statement that we can restrict the balls to having rational radius and centre with real algebraic entries. 
The radius restriction is simple, as we can choose $\mathcal B'$ arbitrarily small.
Hence in particular we can choose the radius to be rational.

For the centre, notice that the initial configuration $\vect d= (2, 2, 0, 0, 0, 2 \pi \ell)$ is not a priori algebraic. 
We can however restrict ourselves to choosing $\ell$ of the form $\frac{q}{\pi}$, with $q \in \mathbb{Q}$. This does not impede our search for lower and upper bounds
 on $L_{\le n_2}(\theta)$: we can define $r,r' \in \mathbb{Q}$ such that 
 $r < \frac{q}{\pi}-0.45 \varepsilon < L_{\le n_2}(\theta) <  \frac{q}{\pi}+ 0.45\varepsilon < r'$, with $r'-r < \varepsilon$. 

 Now, this choice of $\ell=\frac{q}{\pi}$ makes $\vect d= (2, 2, 0, 0, 0, 2 \pi \ell)$ rational. As the linear operator $M^nH$ that transforms the coefficient space to the input space is algebraic, it means that $M^nH\vect d$ is algebraic as well. The normal to $\mathcal B$ at $d$ is rational, thus the normal to $\mathcal{O}=M^nH \mathcal B$ at the algebraic $M^nH\vect d$ is algebraic, and we obtain a centre of $\mathcal B' \subset \mathcal O$ that is algebraic (since the radius is rational).
}
\end{proof}

This completes the proof of Theorem~\ref{th.epsilonrobust}.1.a for closed balls.

\subsection{Case of Open Balls and robust Skolem}
In this subsection, we extend the proof of Theorem \ref{th.epsilonrobust} to show that considering open or closed balls does not make a difference for the Diophantine hardness. Further, there is also no difference whether we consider the robust Skolem problem (0 is avoided), the robust positivity problem (negative numbers are avoided), or the robust strict positivity problem (negative and 0 are avoided). Thus, this establishes Theorem~\ref{th.epsilonrobust}.1.b; and 1.a for open balls.

Let $\ball$ be an open ball and $cl(\ball)$ its topological closure, which is the closed ball consisting of $\ball$ and its surface. 
{\color{black} With the next lemma, we argue that open and closed balls can very often be reasoned about interchangeably.}
Consider the following statements:
\begin{enumerate}
\item Robust positivity holds for the closed ball $cl(\ball)$.
\item Robust positivity holds for the open ball $\ball$.
\item Robust strict positivity holds for the open ball $\ball$.
\item Robust Skolem holds for the open ball $\ball$
\item Robust strict positivity holds for the closed ball $cl(\ball)$
\item Robust Skolem holds for the closed ball $cl(\ball)$
\end{enumerate}

We show that equivalence results between these statements. This allows us to conclude that having open or closed balls does not make a difference for $\mathcal{T}$-Diophantine hardness of Skolem and (strict) positivity. Formally, we have the following.

\begin{lem}
\label{openclose}
(1), (2) and (3) are equivalent. Further, for balls $\ball$ containing at least one initial configuration $\vect d_0$ in its interior that is strictly positive, i.e. $u_n(\vect d_0) > 0$ for all $n$, both (3) and (4) are equivalent and (5) and (6) are equivalent.
\end{lem}

\begin{proof}
(1) implying (2) is trivial.
(2) implies (1): we show the contrapositive. Suppose there exists an initial configuration $\vect d$ on the surface of the ball $\ball$ and an integer $n$ such that $u_n(\vect d) = y < 0$. Recall that $M$ is the companion matrix, and $u_n(\vect d)$ is the first component of $(M^n . d)$, so $u_n(x)$ is a continuous function. Thus, there exists a neighbourhood of $\vect d$, such that for all $\vect d'$ in the neighbourhood, $u_n(\vect d') < y/2 < 0$. This neighbourhood intersects the open ball $\ball$ enclosed by the surface, and picking $d'$ in this intersection shows that Robust Positivity does not hold in the open ball.

(3) implying (2) is trivial.
(2) implies (3): Assume for the sake of contradiction that there is an initial configuration $\vect c'$ in the open ball $\ball$ such that $u_n(\vect c')=0$. Consider any open $O$ around $\vect c'$ entirely in the open ball $\ball$. We have that $\vect c'$ is on hyperplane $H_n$ by definition. That is, there are initial configurations in $O$ on both sides of $H_n$. In particular, there is an initial configuration $\vect c''$ in $O$, hence in $\ball$, with 
$\vect c'' \notin H_n^+$, i.e. $u_n(\vect c'')<0$, a contradiction with $\ball$ being robustly positive.

(3) implies (4) is trivial.
(4) implies (3): We consider the contrapositive: if we have an initial configuration $\vect d_1$ of $\ball$ which is not strictly positive, then $u_n(\vect d_1)\leq 0$ for some $n$, and there is a barycenter $\vect d_2$ between $\vect d_0,\vect d_1$ which satisfies $u_n(\vect d_2)=0$, i.e. negation of (4).
To be more precise, we can choose
$\vect d_2 = \frac{-u_n(\vect d_1)}{u_n(\vect d_1)-u_n(\vect c)} \vect d_0 + \frac{u_n(\vect d_1)}{{u_n(\vect d_1)-u_n(\vect d_0)}} \vect d_1$. 

Now, (5) and (6) are equivalent for balls containing at least one initial configuration $\vect d_0$ that is strictly positive in its interior (same proof as for the equivalence between (3) and (4) above). However, notice that (5,6) are not equivalent with (1,2,3,4) in general.
\end{proof}

We are now ready to prove Theorem \ref{th.epsilonrobust} for open balls $\ball$. 
It suffices to remark that the centre $\vect c$ 
of ${\ball}$ is strictly in the interior of $P_{dom}$, 
and thus it will be eventually strictly positive by Lemma \ref{cone}, that is there exists $n_2>n_1$ such that $u_n(\vect c)> 0$ for all $n>n_2$, and we can choose $\vect d_0= \mathbf{M}^{n_2}\vect c$.
Hence by Lemma \ref{openclose}, robustness (for $n> \max(n_1,n_2)$) of positivity, strict positivity and Skolem are equivalent on $\mathcal B$, and these are equivalent with robust positivity of $cl(\mathcal B)$ which was proved $\mathcal{T}$-Diophantine hard in the previous section.

It remains to prove Theorem \ref{th.epsilonrobust} for
robust Skolem for closed balls ${\mathcal B}$.
For that, it suffices to easily adapt Lemma \ref{l.positive}, replacing $(u_n(\vect d))_{n > n_2}$ positive by strictly positive, and obtain the $\mathcal{T}$-Diophantine hardness for robust strict positivity of closed balls. We again apply Lemma \ref{openclose} ($(5)$ and $(6)$ are equivalent) to obtain hardness for robust Skolem of closed balls.

{\color{black}
\subsection{Case of Robust Ultimate Positivity}
\label{s.lagrange}

We now turn to approximating $L_\infty(\theta)$ using robust ultimate positivity, i.e., to show Theorem~\ref{th.epsilonrobust}.1.c. The idea is similar to approximating $L(\theta)$ that we developed in previous sections.

Let $\ell>0$ and $\varepsilon>0$.
We use $n_2$ and $\psi$ as defined in Lemma \ref{claim:fail}.

\begin{prop}
  \label{prop.closed}
  $\ball$ is (uniformly) ultimately positive implies that 
  $L_\infty(\theta) \geq \ell - \varepsilon$, 
  and 
  $\ball$ is not (uniformly) ultimately positive implies that 
  $L_\infty(\theta) \leq \ell + \varepsilon$.
  \end{prop}

  \begin{proof}
Recall $L_\infty = \frac{1}{2\pi}\liminf_{n>0}n[2\pi n\theta]$.

Assume that $B_\psi$ is (uniformly) ultimately positive.
    Then in particular $\vect d$ is positive.
    Applying Lemma \ref{l.positive}, we obtain that $n[2\pi n\theta] < 2\pi \ell - \varepsilon$ for only finitely many $n$, and hence the limit infimum
    $L_\infty(\theta) \geq \ell - \varepsilon$.

    Now, assume that 
    $\ball$ is not (uniformly) ultimately positive.
    Then by Lemma \ref{claim:fail}, 
    which we can apply as $\psi$ is sufficiently small,
    for infinitely many $n>n_2$, we have that 
    $n[2\pi n \theta] < 2\pi\ell + \varepsilon$.
    Hence the limit infimum $L_\infty(\theta) \leq \ell + \varepsilon$. 
    \end{proof}

This settles the case of closed balls. Now, when we have a uniform constant on ultimate positivity, the case of open balls is equivalent to the case of closed balls, as shown in the following Lemma:

\begin{lem}
  Let $\ball$ be an open ball. Then $cl(\ball)$ is robustly uniformly ultimately positive iff $\ball$ is robustly uniformly ultimately positive.
\end{lem}

\begin{proof}
  One implication is obvious as $\ball \subseteq cl(\ball)$.
  For the other, assume that $\ball$ is robustly uniformly ultimately positive.
  Let $N$ be the uniform constant for ultimate positivity over $\ball$.
  So for all $n>N$, we know that for all $\vect d' \in \ball$, $u_n(\vect d')\geq 0$. Fix a $n>N$.  
  By continuity, we obtain that for all 
  $\vect d' \in cl(\ball)$, we also have $u_n(\vect d')\geq 0$.
  Hence $cl(\ball)$ is robustly uniformly ultimately positive with the same uniform ultimate positive constant.
\end{proof}

This terminates the proof of the last cases of Theorem \ref{th.epsilonrobust}.
Notice that this last Lemma is not true when the constant on ultimate positivity is not uniform over the ball. And indeed, we will show a major complexity difference in the next section.

}

\section{Proof of Theorem \ref{th.robust}}
\label{sec:dec}

We now turn to the proof of Theorem \ref{th.robust}, generalising elements from Section \ref{sec:geom}.

\subsection{Intuitions for the proof of Theorem \ref{th.robust}}
\label{subsec.intuition}

Let $\lrs u$ be a recurrence relation defined by coefficients $\vect a \in \mathbb{Q}^\kappa$. As before, we will consider $\lrs v = (\frac{u_n}{f_n})_{n \in \mathbb{N}}$, for $f_n$ such that the dominant coefficients of $\lrs v$ are of the form $\alpha e^{i n \theta}$.
We will then decompose the exponential solution of $\lrs v$ into two:
a {\em dominant term} $(v_n^{dom})_{n \in \mathbb{N}}$ made of 
coefficients $\alpha e^{i n \theta}$, 
and a {\em residue} $(v_n^{res})_{n \in \mathbb{N}}$ with
$(v_n^{res})_{n \in \mathbb{N}} \underset{n\to +\infty}{\longrightarrow} 0$.
Recall that, denoting by $\vect c^{dom}$ the projection 
of an initial configuration $\vect c$ on dominant space, 
we defined $P_{dom} = \{\vect c \mid \forall n, v_n^{dom}(\vect c^{dom}) \geq 0\}$.

{
\color{black}
The non-negativity of $v_n^{dom}(\vect c)$ for all $n$, and thus membership in $P_{dom}$, is necessary for the Ultimate Positivity of the LRS initialised by $\vect c$. However, the decidability of this prerequisite is unclear from the above formulation of $P_{dom}$. We define a function $\dominant(\vect c, \vect t)$, whose second argument $\vect t$ comes from a continuous domain, namely the Masser Torus $T$ \cite{Masser}. This torus $T$ is compact, and Lemma \ref{lem:pdomtorus} establishes an important closure property: $P_{dom}$ can equivalently be defined as the set of points $\vect c$ for which $\mu(\vect c) = \min_{\vect t \in T} \dominant(\vect c, \vect t)$ is non-negative. This definition is more accessible: Renegar's theorem \cite{Renegar} states that $\mu(\vect c)$ can be computed explicitly, which we will use in some cases but manage to avoid in others, as detailed below.
}

Before giving the formal proofs, we provide our intuitions for all 4 decidable cases. We start with the simplest case $\exists$-robust ultimate positivity, then move to robust non-uniform ultimate positivity, then $\exists$-robust positivity and finally $\exists$-robust Skolem.

\begin{enumerate}
\item For $\exists$-robust uniform ultimate positivity, we consider different possibilities for $\mu=\mu(\vect c_0)$, and show the following in Proposition \ref{prop:critical}:
\begin{itemize}
   \item If $\mu<0$, then $\vect c_0$ is not in $P_{dom}$, 
     that is $u_n(\vect c_0)<0$ for an infinite number of indices $n \in \mathbb{N}$, so $\vect c_0$ cannot be ultimately positive and we are done.
  \item If $\mu=0$, then $\vect c_0$ is on the surface of $P_{dom}$, so it is arbitrarily close to point not in $P_{dom}$, so $\vect c_0$ cannot be {\em robustly} ultimately positive.
  \item The last case $\mu>0$, means that $\vect c_0$ is in the interior of $P_{dom}$, and so in particular there exists a closed ball $\cal B$ around $\vect c_0$ entirely in the interior of $P_{dom}$. In particular, for some $N$ large enough, $\cal B$ is in the intersection of all the half-space ${\mathcal H}_n, n>N$, and thus $\vect c_0$ is robustly uniformly ultimately positive for the uniform bound $N$.
    \end{itemize} 
Thus, we have reduced deciding $\exists$-robust uniform ultimate positivity to checking the sign of $\mu(\vect c_0)$. As mentioned earlier we can explicitly compute $\mu(\vect c_0)$ using Renegar's result~\cite{Renegar} and see whether it is positive or not. In this case, we actually do not need to compute $\mu(\vect c_0)$ explicitly, only test whether it is strictly positive (since that is the only case where we can be $\exists$-robust ultimately positive), which can be formulated as an FO formula over Reals.

\item The next case is robust {\em non-uniform} ultimate positivity for an {\em open} ball $\mathcal B$ of algebraic radius $\psi$, centred in $\vect c_0$. We show in Proposition~\ref{prop.open} that $\mathcal B$  is robustly {\em non-uniformly} ultimately positive iff $\mathcal B \subseteq P_{dom}$, i.e. iff $\mu(\vect c)>0$ (the inequality is strict as $\mathcal B$ is open and $P_{dom}$ is closed) for all $\vect c \in \mathcal B$, which can be effectively tested in $\pspace$ using a FO formula over reals.

 \item For $\exists$-robust positivity, we again do a case analysis according to the signs of $\mu=\mu(\vect c_0)$, by observing that:
  \begin{itemize}
    \item if $\mu<0$, that is $\vect c_0 \notin P_{dom}$, then there exists a $n$ such that the LRS initialised by $\vect c_0$ is negative, so $(u_n(\vect c_0))_{n \in \mathbb{N}}$ is not robustly positive;
    
    \item if $\mu=0$, that is $\vect c_0$ is at the surface of $P_{dom}$, then there exists a configuration arbitrarily close to $\vect c$ such that the LRS from that configuration is negative, i.e. $(u_n(\vect c_0))_{n \in \mathbb{N}}$ is not robustly positive either;
    
    \item if $\mu>0$, that is $\vect c_0 \in P_{dom}$ and not on the surface, then as the residue has negligible contribution to $\lrs v$ for large $n$, we show that the LRS will ultimately avoid negative numbers beyond a threshold index $n_{thr}$ depending in the exact value of $\mu$, which can be computed using Renegar's result \cite{Renegar}.
  \end{itemize}

   Having assured ourselves of the long run behaviour, it suffices to check the value of the LRS up to $n_{thr}$, where the residue can have significant contribution, to see whether the LRS is strictly positive, in which case it satisfies robust positivity.   

   \item The last case is for $\exists$-robust Skolem. {\color{black} Analogous to robust positivity problem that sought to avoid negative numbers and hence required $v_n^{dom}$ to be always positive, here we seek to avoid $0$ and thus require $v_n^{dom}$ to be non-zero, i.e. have strictly positive absolute value. We thus concern ourselves with $\nu = \min_{\vect t \in T} |\dominant(\vect c, \vect t)|$, where $T$ is the continuous, compact torus.} We split our analysis based on the possible values of $\nu$, which we show in Proposition \ref{prop:Renegar} can be computed effectively:
  \begin{itemize}
  \item $\nu>0$. Then as the residue has negligible contribution to $\lrs v$ for large $n$, we show that the LRS will ultimately avoid zero beyond a threshold index $n_{thr}$. Having assured ourselves of the long run behaviour, it suffices to check the value of the LRS up to $n_{thr}$, where the residue can have significant contribution, to see whether the LRS satisfies robust Skolem.
  \item $\nu=0$. Then we show in Proposition~\ref{prop:nuprop} that   the LRS does not satisfy robust Skolem: no matter how small we pick a neighbourhood around $\vect c_0$, there will always exist a $\vect c$ in that neighbourhood that hits zero at some iteration.
  \end{itemize}

\end{enumerate}

The rest of this section formalises the above intuitions to prove Theorem~\ref{th.robust}.

\subsection{Masser's Torus and relation with $P_{dom}$.}

We first define the normalised exponential polynomial solution $\lrs v$:
\begin{defi}
\label{defn:normalLRS}
 Let $\lrs u$ be an LRS of general term $u_n(\vect c) = \sum_{i=1}^r\sum_{j=0}^{m_r-1}p_{ij}n^j\gamma_i^n$, with $\rho$ being the modulus of the dominant roots and $m+1$ the maximal multiplicity of a dominant root. Define $v_n(\vect c) = \frac{u_n(\vect c)}{n^m\rho^n}$ for $n>0$, and $v_0(\vect c) = u_0(\mathbf{c})$.
 \end{defi}

We call every term of $v_n$ which converges towards 0 as $n$ tends towards infinity residual, while the other terms, of the form $\alpha e^{i \theta}$ are dominant. We denote $ \{\theta_j \mid j = 1, \ldots, k\}$ the set of $\theta$ in dominant terms,  and
$\alpha_j(\vect c)$ the associated coefficient. We denote the total contribution from the dominant terms of the form $\alpha_j(\vect c)e^{in\theta_j}$ by $v_n^{dom}(\vect c)$. We have that:
$$v^{dom}_n(\vect c) = \sum \alpha_j(\vect c) e^{i n\theta_j}\quad \text{and}\quad v_n^{res}(\vect c) = v_n(\vect c)-v_n^{dom}(\vect c)= \OO(\frac{1}{n}) \rightarrow_{n \rightarrow \infty} 0.$$

As we explained in Section \ref{subsec.intuition}, knowing 
whether $(v_n^{dom})_{n \in \mathbb{N}} >0$ for all $n \in \mathbb{N}$
is crucial in order to solve $\exists$-robust Skolem, positivity, and (uniform or not) ultimate positivity. In Section~\ref{sec:geom} and~\ref{sec:hardness}, we dealt with hardness via an example which had 3 dominant roots, and it was rather simple to determine the min/max value (computed in the proof of Lemma \ref{cone}). The general case is not so easy however. We follow the method defined in \cite[Theorem 4]{ouaknine2014positivity}.
Towards this goal, define $\vect s^n=(e^{i n \theta_1}, \ldots, e^{i n \theta_k})$ for all $n \in \mathbb{N}$. Computing the range of $\vect s^n$ over $\mathbb{N}$ is not simple.
One idea is to perform a (continuous) relaxation, as provided by (a weak version of) Kronecker's Theorem:

\begin{thm}[Kronecker \cite{Kronecker}]
  \label{thm:kronecker}
  Let $\theta_1, ... , \theta_k, \varphi_1, ... , \varphi_k \in [0, 2\pi)$
such that for every tuple $(\lambda_1,...\lambda_k)$ of integers with 
    $e^{i (\lambda_1 \theta_1 + \cdots +  \lambda_k \theta_k)}=1$, 
  we have $e^{i (\lambda_1 \varphi_1 + \cdots + \lambda_k \varphi_k)}=1$.

  Then for any $\epsilon > 0$, there exists an arbitrarily large $n$ such that for all 
    $1 \le j \le k$ we have $|e^{i n \theta_j} - e^{i \varphi_j}| \le \epsilon$.
  \end{thm}

  Intuitively, one can replace 
  $\{ \vect s^n = (e^{i n \theta_1}, \ldots, e^{i n \theta_k}) \mid n \in \mathbb{N} \}$ by a set $T$ as long as 
  \begin{enumerate}
    \item  for all $n$,  $\vect s^n = (e^{i n \theta_1}, \ldots, e^{i n \theta_k}) \in T$
and,
\item  for all $(e^{i \varphi_1}, \ldots, e^{i \varphi_k}) \in T$, we have
$e^{i (\lambda_1 \varphi_1 + \cdots + \lambda_k \varphi_k)}=1$ for 
every tuple $(\lambda_1,...\lambda_k)$ of integers such that 
$e^{i (\lambda_1 \theta_1 + \cdots +  \lambda_k \theta_k)}=1$.
  \end{enumerate}

  The first item ensures that $T$ is a relaxation of 
  $\{ \vect s^n \mid n \in \mathbb{N} \}$, while the second ensures that 
  $\{ \vect s^n \mid n \in \mathbb{N} \}$ is dense in $T$ by Kronecker's Theorem \ref{thm:kronecker}. 
  To define such a set $T$, we have to extract the integer multiplicative relations between the roots $e^{i \theta_1}, \ldots, e^{i \theta_k}$, to fulfil the second requirement.
  A tedious but now rather classical way to obtain these relations 
  is by invoking a deep result from Masser \cite{Masser}:
  
  \begin{thm}[Masser \cite{Masser}]
  \label{thm:abelian}
  Let $K \in \mathbb{N}$ be fixed. Let $e^{i \theta_1},...,e^{i \theta_k}$ be complex algebraic numbers of unit modulus, given as input. Consider the free abelian group $L$ defined by $L = \{(\lambda_1, \ldots ,\lambda_k) \in \mathbb{Z}^k: 
  e^{i (\lambda_1 \theta_1 + \cdots +  \lambda_k \theta_k)} = 1 \}$. 
  The group $L$ has a finite generator set $\{ \mathbf{l_1}, \ldots, \mathbf{l_p}\} \subset \mathbb{Z}^k$ with $p \le k$. Each entry in the generator set is polynomially bounded in 
  $\size(e^{i \theta_1}), \ldots, \size(e^{i \theta_k})$. The generator set can be computed in polynomial space. Further, if the decision problem guarantees that $k$ is at most $K$, the generator set can be computed in polynomial time.
  \end{thm}

  Using the generator set of $L$ provided by Theorem \ref{thm:abelian},
we can define a set $T$ as a Torus as follows:
$$T = \{ \vect t = (e^{i \varphi_1},\ldots, e^{i \varphi_k}) \mid \forall \, \vect l=(\ell_1, \ldots, \ell_k) \in L, e^{i (\ell_1 \varphi_1 + \cdots + \ell_k \varphi_k)} = 1\}.$$

By definition of $T$, we have $\vect s^n \in T$ for all $n \in \mathbb{N}$, so the first requirement is fulfilled.
For the second requirement, let $(e^{i \varphi_1}, \ldots, e^{i \varphi_k}) \in T$
and $(\lambda_1,...\lambda_k) \in \mathbb{Z}^k$ such that 
$e^{i (\lambda_1 \theta_1 + \cdots +  \lambda_k \theta_k)}=1$.
By the definition of a generator set, there exist 
$\alpha_1,\ldots, \alpha_p \in \mathbb{Z}^k$ with 
$(\lambda_1,...\lambda_k) = \sum_{1}^p \alpha_i \vect l_i$.
In particular, 
$e^{i (\lambda_1 \varphi_1 + \cdots +  \lambda_k \varphi_k)}=
\prod_{\vect l=(\ell_1,\ldots, \ell_k) \in L} [e^{i(\ell_1 \varphi_1 + \ldots + \ell_k \varphi_k)}]^{\alpha_i} = \prod_{\vect l=(\ell_1,\ldots, \ell_k) \in L} 1^{\alpha_i} = 1$, and the second requirement on $T$ from above is also fulfilled.

Notice that the Torus $T$ is \emph{independent of the initial configuration $\vect c$}. 

Denoting, for every initial configuration $\vect c$ and element of the torus $\vect t \in T$, $\dominant(\vect c, \vect t) = \sum_j \alpha_{j}(\vect c) t_j$, we obtain that 
  $v_n^{dom}(\vect c)=\dominant(\vect c, \vect s^n)$ for all $n$ and all $\vect c$. 
  We have that $v_n^{dom}(\vect c) \in \{\dominant(\vect c, \vect t) \mid \vect t \in T\}$.
  The second statement translates into:
  \begin{equation}
    \label{eq100}
    \forall \vect t \in T, \varepsilon > 0, \exists n, \text{ with } 
    \forall \vect c, |v_n^{dom}(\vect c)-\dominant(\vect c, \vect t)| \leq \varepsilon.
  \end{equation}

Using the above, we can relate $T$ with $P_{dom}$:

\begin{lem}
  \label{lem:pdomtorus}
$$P_{dom} = \{\vect c \mid \forall t \in T, \dominant(\vect c, \vect t) \geq 0\}.$$
\end{lem}

\begin{proof}
  Let us denote $P'_{dom} = \{\vect c \mid \forall t \in T, \dominant(\vect c, \vect t) \geq 0\}$. Remember that the definition is 
$P_{dom} = \{\vect c \mid \forall n \in \mathbb N, v_n^{dom}(\vect c) \geq 0\}$.
We now prove that $P'_{dom}=P_{dom}$.
As for all $n$ and all $\vect c$, we have $v_n^{dom}(\vect c)=\dominant(\vect c, \vect s^n)$, we have that  $P'_{dom} \subseteq P_{dom}$ directly.

We now prove that $P_{dom} \subseteq P'_{dom}$. Let $\vect c \in P_{dom}$, i.e.
$\forall n \in \mathbb N, v_n^{dom}(\vect c) \geq 0$. 
Assume by contradiction that $\vect c \notin P'_{dom}$.
Hence there exists $\vect t \in T$ with $\dominant(\vect t,\vect c) <0$.
Denote $\varepsilon = - \frac{\dominant(\vect t,\vect c)}{2} > 0$.
Using (\ref{eq100}), we obtain an $n$ with $v_n^{dom}(\vect c) < - \frac{\varepsilon}{2} < 0$, a contradiction with $\vect c \in P_{dom}$.
\end{proof}

\subsection{Relation between sign of \texorpdfstring{$\mu(\vect c)$}{μ(c)} and (robust) positivity.}

Let $\vect c$ be an initial configuration and let $T$ be Masser's Torus as described above. Then, we define 
\begin{align}\mu(\vect c) = \min_{\vect t \in T} \dominant(\vect c, \vect t). \label{eq:mu}\end{align}

We now state the crucial proposition that will allow us to 
relate the sign of $\mu(\vect c)$ with the position of $\vect c$ with respect to 
$P_{dom}$, which will be used to characterise $\exists$-robust positivity and ultimate positivity around $\vect c$.

\begin{prop}
\label{prop:critical} \hfill
\begin{enumerate}
  \item If $\mu(\vect c) < 0$, then (i) $\vect c \notin P_{dom}$ and (ii) there exist an infinite number of $n$ such that 
$v_n(\mathbf{c}) < 0$.

\item If $\mu(\vect c) = 0$,
then (i) $\vect c$ is on the surface of $P_{dom}$ and (ii) $\forall \varepsilon>0$, 
$\exists \vect c_\varepsilon$ such that 
$|\vect c - \vect c_\varepsilon|\leq \varepsilon$
and an infinite number of $n$ with $v_n(\mathbf{c}_{\varepsilon}) < 0$.
\item If $\mu(\vect c)>0$, 
then (i) $\vect c$ is in the interior of $P_{dom}$ and 
for all $\varepsilon >0$, (ii) there exists $N$, such that for all 
 $\vect c_\varepsilon$ with $|\vect c - \vect c_\varepsilon |\leq \varepsilon$
 and all $n>N$, we have $v_n (\mathbf{c}_\varepsilon) > 0$. 
  \end{enumerate}

\end{prop}

\begin{proof}

  First, we will prove a useful claim, relating $v_n(\vect c)$ and the distance from $\vect c$ to hyperplane $H_n$:
  For every $n$, let $distance(\vect c,H_n)$ be the distance between an initial configuration $\vect c$ and the hyperplane  $H_n = \{\vect c' \mid v_n(\vect c')=0\} = \{\vect c' \mid u_n(\vect c')=0\}$.

  \begin{clm}
  \label{l.distance}
  There exists $C$ such that 
  for all $n$, $distance(\vect c,H_n) \leq C \cdot |v_n(\vect c)|$.
  \end{clm}
  
  \begin{proof}
    Let $n \in \mathbb{N}$. 
    We have $distance(\vect c,H_n)= \frac{|u_n(\vect c)|}{||\vect y||}$
    for $\vect y$ the first row of $\M^n$ by basic geometry.
    Let $H$ be the transformation matrix between 
    the basis of initial configurations and
    the basis of the exponential polynomial
    solution of $\lrs u$.
    Let $\vect x=(x_1, \ldots, x_\kappa)$ with $x_i=n^k \rho^n_j$
    so that to cover every root $\rho_j$ and multiplicities $k = 1, \ldots, m_j$.
    We have $u_n(\vect c) = \vect y \cdot \vect c = \vect x \cdot (H \cdot \vect c)$ for all initial configurations $\vect c$, i.e., 
    $\vect y = \vect x \cdot H$.
    That is, there exists a constant $D>0$ depending upon $H$ with 
    $||\vect y|| \geq D n^m \rho^n$ for 
    $\rho$ the modulus of a dominant root and $m+1$ the highest multiplicity of a root of modulus $\rho$.
    We obtain 
    $distance(\vect c,H_n) \leq  \frac{|u_n(\vect c)|}{D n^m \rho^n} = \frac{|v_n(\vect c)|}{D}$.
    \end{proof}
    
  Now we prove the relationship between the sign of $\mu(\vect c)$ and the position wrt $P_{dom}$.
  \begin{itemize}
    \item If $\mu(\vect c) < 0$, then let $\varepsilon = - \mu(\vect c) > 0$.
Using Equation~(\ref{eq100}), taking $t \in T$ realising the minimum value $\mu(\vect c)$,
we have an $n$ such that $v_n^{dom}(\vect c) - \dominant(\vect c,\vect t) \leq \frac{\varepsilon}{2}$. In particular, $v_n^{dom}(\vect c) \leq - \frac{\varepsilon}{2} <0$ and $\vect c \notin P_{dom}$.
\item If $\mu(\vect c) > 0$, then in particular for all $n$, 
$v_n^{dom}(\vect c) = \dominant(\vect c,s^n) \geq \mu(\vect c) > 0$, meaning that 
  $\vect c \in P_{dom}$, in its interior.
  \item The last case is $\mu(\vect c) = 0$. 
Then by the previous reasoning, $\vect c \in P_{dom}$, but not necessarily in the interior. To see that it is on the surface of $P_{dom}$,
it suffices to use Equation~(\ref{eq100}) again. By contradiction, assume that it is not the surface of $P_{dom}$. Hence there exists a distance $\varepsilon>0$ to the surface, hence at least 
$\varepsilon$ away from all hyperplanes $H_{n}$.
Taking $t \in T$ realising the minimum value $\mu(\vect c)$,
we have an $n$ such that $|v_n^{dom}(\vect c) - \dominant(\vect c,\vect t)| \leq \frac{\varepsilon}{2C}$. In particular, $v_n^{dom}(\vect c) \leq \frac{\varepsilon}{2C}$,
a contradiction with $distance(\vect c,H_n) > \varepsilon$ using Claim \ref{l.distance}.
\end{itemize}This completes the proof of $1(i)$, $2(i)$ and $3(i)$.

\medskip

We now turn to the existence of counterexamples of positivity  
in the neighbourhood of $\vect c$ or of a neighbourhood of $\vect c$ entirely positive, 
i.e., proof of $1(ii)$, $2(ii)$ and $3(ii)$. For this, we use Claim \ref{l.distance} in the following way:  if for all $\alpha >0$,  there exist an infinite number of $n_\alpha$ with $|v_{n_\alpha}(\vect c)| < \alpha$, then there exists an infinite number of $n$ with $distance(\vect c,H_n)<\varepsilon$ 
(choose $n= n_\alpha$ for $\alpha=\frac \varepsilon {2C})$. This means that for any neighbourhood we pick, there will be a violation of positivity in it for infinitely many $n$.
\begin{itemize}
\item Let $\mu=0$. We prove that for all $\alpha>0$, we have a $n_\alpha$ such that 
$|v_{n_\alpha}(\vect c)| \leq \alpha$, which suffices by Claim  \ref{l.distance}.  
Let $\alpha>0$ arbitrarily small, and let $N$ such that for all $n>N$, 
  $|v_n^{res}(\vect c)| \leq \frac{\alpha}{2}$. This $N$ exists  as $v_n^{res}(\vect c) \rightarrow_{n \rightarrow \infty} 0$. 
  By Equation~(\ref{eq100}), as $\mu=0$, there exists an (infinite number as $n$ can be chosen arbitrarily large of) $n_{\alpha}>N$ with   $|v^{dom}_{n_{\alpha}}(\vect c)|< \frac{\alpha}{2}$.  Thus $|v_{n_{\alpha}}(\vect c)| \leq |v_{n_{\alpha}}^{dom}(\vect c)| + |v_{n_{\alpha}}^{res}(\vect c)| \leq \alpha$, thus proving 2(ii).
\item Let $\mu <0$. We still have  $|v_n^{res}(\vect c)| \leq \frac{\mu}{2}$
for all $n>N$. Again by Equation~\ref{eq100}, we have an infinite number of $n>N$ such that $v_n^{dom}(\vect c) < \frac{\mu}{2}$, and we get  $v_n(\vect c)=v_n^{dom}(\vect c) + v_n^{res}(\vect c) < \mu$, and thus 1(ii) follows.
\item The last statement is with $\mu(\vect c)>0$. We have that $\vect c$ is guaranteed to be $\varepsilon$ away from all hyperplanes $H_n$ for sufficiently large $n$ (i.e.\ $\ge N$). Hence considering the ball $\mathcal B$ of  radius $\frac{\varepsilon}{2}$ and of centre $\vect c$,  all the initial configurations $\vect c_{\varepsilon} \in \mathcal B$ are at distance at least $\frac{\varepsilon}{2}$ to all hyperplanes $H_n$, and in particular we also have $v_n(\vect c_{\varepsilon})>0$. \qedhere
\end{itemize}
\end{proof}

\subsection{Deciding \texorpdfstring{$\exists$}{∃}-robust ultimate positivity and \newline robust non-uniform ultimate positivity for open balls}

We are now ready to provide the algorithms and proof for the decidable cases of 
robust ultimate positivity. First, we consider $\exists$-robust ultimate positivity. For this case, there is no difference on uniformity of open/closed ball. 

\begin{prop}
  \label{prop.ultimate}  
  Let  $\vect c_0$ be an initial configuration.   $\vect c_0$  is $\exists$-robustly ultimately positive iff   $\mu(\vect c_0)>0$. Further, this condition can be checked in $\pspace$.
\end{prop}

\begin{proof}
We adapt a similar analysis as the one that appeared in \cite{ouaknine2014positivity}.
In one direction, if $\mu(\vect c_0)>0$, then by Proposition \ref{prop:critical} 3., we infer that $\vect c_0$  is $\exists$-robust ultimately positive. In the other direction, suppose $\mu(\vect c_0) \leq 0$, then by Proposition \ref{prop:critical} 1. or 2. we obtain that for all $\epsilon$-neighbourhoods of $\vect c_0$ there are an infinite number of $n$ such that $v_n(\vect c_\varepsilon)<0$, i.e.,  $\vect c_0$  is not $\exists$-robust ultimately positive.

For the complexity, we observe that to check whether $\mu(\vect c_0)>0$, its complement can be checked by a call to an oracle for existential/universal first order theory of reals (which is decidable in $\pspace$). That is, $\vect c_0$ is $\exists$-robust ultimately positive if the following first order (universal) theory of reals sentence holds:
\begin{align}\label{eq:foupos}
  \forall \vect t \in T, \dominant(\vect c_0, \vect t) > 0.
\end{align}

In the above, we briefly remark that $T$ is compact, and hence the minimum value of a dominant is well defined, should one need to compute it. The only difficulty is to express $\forall \vect t \in T$ as a universally quantified formula of the first order theory of reals. For that, it suffices to remark that 
$(a_1+i b_1, \ldots, a_k +i b_k) \in T$ iff 
$||a_i+i b_i||^2=1$ for all $i \leq k$ and 
$\forall \vect l=(\ell_1, \ldots, \ell_k)$ in the generator set, 
where the complex part is $Im((a_1+i b_1)^{\ell_1} \cdots (a_k +i b_k)^{\ell_k})=0$
and the real part is
$Re((a_1+i b_1)^{\ell_1} \cdots (a_k +i b_k)^{\ell_k})=1$. 
Notice that  $a_i,b_i \in \mathbb{R}$ are real variables, and $(\ell_i)_{i \leq k}$ are fixed integers completely determined by the input, with magnitude polynomial in the size of the input. We can assume without loss of generality that $\ell_i \in \mathbb{N}$ for all $i$, 
as $(a_i+i b_i)^{-\ell}= (a_i-i b_i)^{\ell}$ since 
the modulus $||a_i+i b_i||=1$. This makes the size of the obtained universal sentence polynomial in the size of the input, and we can indeed determine its truth in $\pspace$.
\end{proof}

\begin{algorithm}[t!]
  \caption{to check robust non-uniform ultimate positivity for an open ball $\mathcal B$ and
  to check $\exists$-robust ultimate positivity of $\vect c_0$.}
  \label{algoU}
  \SetAlgoLined
  \DontPrintSemicolon
  \KwIn{Companion matrix $\M \in \mathbb{Q}^{\kappa \times \kappa}$
  of $\lrs{u}$, initial configuration $\vect c_0 \in \mathbb{Q}^{\kappa}$, open ball $\mathcal{B}$ with centre $\vect c_0'$ and radius $\psi$}
  
  Compute $\{\gamma_j\}_j \gets$ eigenvalues of $\M$, \quad $\rho \gets \max_j |\gamma_j|$, \quad $k \gets |\{\gamma_i/\rho ~|~ |\gamma_i| = \rho\}|$, 
  $\{e^{i \theta_j}\}_{j=1}^k \gets \{\gamma_i/\rho ~|~ |\gamma_i| = \rho\}$ \;
  
  Determine $T$ Torus obtained by applying Masser's result (Theorem \ref{thm:abelian}) to $\{\theta_j\}_{j=1}^k$ \;
  
  \If{$\forall \vect t \in T, \dominant(\vect c_0, \vect t) > 0$ (check using Equation~\ref{eq:foupos})}
  {\Return $\vect c_0$ is $\exists$-robust ultimate positive}
  
  \If{ $\forall \vect d\in \mathbb{R}^\kappa, \forall \vect t \in T, ||\vect d || \leq \psi \rightarrow ~\dominant(\vect c'_0+ \vect d, \vect t) \geq 0$ (check using Equation~\ref{eq:forupos})}
  {\Return $\mathcal{B}$ is robust non-uniform ultimate positive}
  
  \end{algorithm}

\medskip

Next, we turn to the decidability of robust {\em non-uniformly} ultimate positivity for {\em open} balls, which is slightly more complex.

\begin{prop}
  \label{prop.open}  
  Let  $\mathcal B$ an {\em open} ball with centre $\vect c_0$ and radius a real algebraic number $\psi$.   $\mathcal B$  is robustly {\em non-uniformly} ultimately positive iff ${\mathcal B} \subseteq P_{dom}$. Further, this condition can be checked in $\pspace$.
\end{prop}
 \begin{proof}
In one direction, if ${\mathcal B} \subseteq P_{dom}$, then for all  $\vect c \in \mathcal B$, we have $\vect c \in P_{dom}$. In fact, we obtain that $\vect c$ is in the relative interior of the set $P_{dom}$ since $\mathcal B$ is open. By Proposition~\ref{prop:critical}, this implies that for any such  $\vect c \in {\mathcal B}$, $\mu(\vect c)>0$. This follows since if $\mu(\vect c) \leq 0$, that would imply $\vect c$ is either on the surface of $P_{dom}$ or not in $P_{dom}$ at all, which would contradict the assumption. Now, let $N_{\vect c}$ such that $v^{res}_n(\vect c)$ is negligible wrt $\mu(\vect c)$. Then $u_n(\vect c) >0$ for all $n > N_{\vect c}$. That is, ${\mathcal B}$  is robustly (non uniformly) ultimately positive.

In the other direction, if ${\mathcal B} \not \subseteq P_{dom}$, then consider a configuration $\vect c \in  {\mathcal B} \setminus P_{dom}$. That is, $\mu(\vect c)<0$ (again by contradiction as $\mu(\vect c) \geq 0$ would mean $\vect c$ is in $P_{dom}$). Applying Proposition \ref{prop:critical}, there exist an infinite number of $n$ such that $u_n(\vect c) < 0$, that is, ${\mathcal B}$ is not robustly ultimately positive.

Finally, we show how to effectively test whether ${\mathcal B} \subseteq P_{dom}$.  Recall that $\vect c_0$ is the centre of $\mathcal B$ and $\psi$ its radius. Then, it suffices to solve the following first order theory of reals sentence, using the effective first order definability of Torus $T$ as before. 
\begin{align}
  \forall \vect d\in \mathbb{R}^\kappa, \forall \vect t \in T, ||\vect d || \leq \psi \rightarrow ~\dominant(\vect c_0+ \vect d, \vect t) \geq 0.
  \label{eq:forupos}\end{align}
We remark in particular that this sentence is part of the universal fragment, and its truth can thus be decided in $\pspace$, as required.
 \end{proof}

\subsection{Decidability for \texorpdfstring{$\exists$}{∃}-robust positivity and \texorpdfstring{$\exists$}{∃}-robust Skolem}

Finally, we consider decidability for $\exists$-robust Skolem and positivity as stated in Theorem \ref{th.robust}, using Proposition \ref{prop:critical}. Unlike for ultimate positivity, where we do not need to resort to~\cite{Renegar}, here we will have to compute $\mu(\vect c_0)$ explicitly, which can be done using Renegar's result. More precisely, for $\exists$-robust positivity and Skolem, we need to check a number of first steps. This number can be obtained effectively given $\min_{\vect t \in T} \dominant(\vect c_0, \vect t)$, which is precisely was is being computed by Renegar's result \cite{Renegar}. 

The procedure for deciding $\exists$-robust positivity is given in Algorithm~\ref{algoP}. 
The rationale for its correctness is as follows:
first, we compute $\mu \gets \min_{\vect t \in T}|\dominant(\vect c_0, \vect t)|$ using 
\cite{Renegar}, %
for $\vect c_0$ the initial configuration around which we are looking for a neighbourhood. If $\mu\leq 0$, then we declare $\exists$-robust positivity does not hold. Otherwise, we compute $N_{thrpos}$ such that 
$|v_n^{res}(\vect c_0)| < \frac{\mu}{2}$ for all $n> N_{thrpos}$. 
Such a $N_{thrpos}$ exists as $|v_n^{res}(\vect c_0)|$ tends to $0$ for $n$ tending towards $\infty$.

Then, we check if $v_{n}(\vect c_0) \leq 0$ for some $n \leq N_{thrpos}$.
If yes, then $\exists$-robust positivity does not hold as $c_0$ 
will be arbitrarily close to initial configuration $c$ with 
$v_{n}(\vect c) < 0$.
Otherwise, $\exists$-robust positivity holds, 
because we know that after $N_{thrpos}$ steps, 
$|v_n^{res}(\vect c_0)| < \frac{\mu}{2}$, and thus 
$v_n(\vect c)=v_n^{dom}(\vect c_0) + v_n^{res}(\vect c_0)  > \mu - \frac{\mu}{2} \geq \frac{\mu}{2}$. This lower bound ensures that there is a neighbourhood around $c_0$ which remains positive.

\begin{algorithm}[t!]
\caption{to check $\exists$-robust positivity of initial configuration $\vect c_0$}
\label{algoP}
\SetAlgoLined
\DontPrintSemicolon
\KwIn{Companion matrix $\M \in \mathbb{Q}^{\kappa \times \kappa}$
of $\lrs{u}$ and initial configuration $\vect c_0 \in \mathbb{Q}^{\kappa}$}

Compute $\{\gamma_j\}_j \gets$ eigenvalues of $\M$, \quad $\rho \gets \max_j |\gamma_j|$, \quad $k \gets |\{\gamma_i/\rho ~|~ |\gamma_i| = \rho\}|$, 
$\{e^{i \theta_j}\}_{j=1}^k \gets \{\gamma_i/\rho ~|~ |\gamma_i| = \rho\}$ \;

Determine $T$ Torus obtained by applying Masser's result (Theorem \ref{thm:abelian}) to $\{\theta_j\}_{j=1}^k$ \;

$\mu \gets \min_{\vect t \in T}\dominant(\vect c_0, \vect t)$ (\cite{Renegar}) \;

\eIf{$\mu \leq 0$}
{\Return NO (Proposition \ref{prop:critical})}
{
Compute $N_{thrpos}$ such that $|v_n^{res}(\vect c_0)| < \frac{\mu}{2}$ for all $n> N_{thrpos}$\;

\ForEach{$n \in \{0, 1, \dots, N_{thrpos}\}$}
{
\If{$\mathbf{M}^n \vect c_0 \leq 0$}
{
\Return NO \;
}
}
\Return YES \;
}
\end{algorithm}

{\color{black} Decidability for arbitrary real algebraic input is clear because each step of the algorithm is decidable (resp.\ computable). We now argue about the complexity when the input is rational, and the a priori bound on the order of the LRS is hard-wired into the decision problem. The assumption on the bounded order ensures both $\mu$ and $1/\mu = 2^{s^{\Oone}}$ are bounded by \cite{Renegar}. We have $N_{thrpos} = 2^{s^{\Oone}}$ because $v^{res}_n(\vect c_0) = \OO(\frac{1}{n})$. This is the number of iterates we have to explicitly check, which gives the $\pspace$ complexity for rational input. 
This is because our strategy involves guessing the index of violation in $\mathsf{coNP}$, constructing an integer straight line program that computes the term at the guessed step via iterated squaring, and then using a $\mathsf{PosSLP}$ oracle to check whether the centre of the neighbourhood is strictly positive at the guessed iterate. $\mathsf{PosSLP}$ was famously shown by Allender et.\ al.\ to be in $\pspace$ \cite{allender}. A rational LRS can easily be scaled up to an integer LRS\footnotemark, that preserves the signs of each term, and enables it to be cast as a division-free integer straight line program.
\footnotetext{To scale the characteristic roots by $\rho$, replace the characteristic polynomial $X^\kappa - a_{k-1}X^{\kappa-1} \dots - a_0$ by $X^\kappa - \rho a_{\kappa-1}X^{\kappa-1} - \rho^\kappa a_0$. Similarly, replace the initial terms by $u_0, \rho u_1, \dots, \rho^{\kappa-1}u_{\kappa-1}$. This results in replacing the sequence $(u_n)_n$ by the integer $(\rho^n u_n)_n$, for an appropriate choice of $\rho$, e.g. least common multiple of denominators.}
}

\medskip

We now turn to deciding Robust $\exists$-Skolem. While for $\exists$-Robust positivity we considered $\mu$, for Skolem we will need its absolute value counterpart. Namely, we define:
 \begin{align} \nu = \min_{\vect t \in T} |\dominant(\vect c_0, \vect t)|. \label{eq:nu}\end{align}

We now show how to adapt Renegar's result~\cite[Theorems 1.1 and 1.2]{Renegar} to effectively compute $\nu$ when the order of LRS is bounded a priori.

{\color{black}
\begin{thm}[Tarski \cite{tarski}]
For every formula $\tau(\vect y)$ in the First Order Theory of the Reals, we can compute an equivalent quantifier-free formula $\chi(\vect y)$.
\end{thm}

In specific scenarios, this computation can be considered efficient.

\begin{thm}[Renegar]
\label{theorem:renegar}
Let $M \in \mathbb{N}$ be fixed. Let $\tau(\mathbf{y})$ be a First Order formula with free variables $\mathbf{y}$, interpreted over the Theory of the Reals. Assume that the total number of free and bound variables in $\tau(\mathbf{y})$ is bounded by $M$. Denote the maximum degree of the polynomials in $\tau(\mathbf{y})$ by $d$ and the number of atomic predicates in $\tau(\mathbf{y})$ by $n$. Then there is a procedure which computes an equivalent quantifier-free formula
\begin{equation*}
\chi(\mathbf{y}) = \bigvee_{i=1}^I \bigwedge_{j=1}^{J_i} h_{i, j}(\mathbf{y}) \sim_{i, j} 0
\end{equation*}
in disjunctive normal form, where each $\sim_{i, j}$ is either $>$ or $=$, with the following properties:
\begin{enumerate}
\item Each of the $i$ and $J_i$ (for $1 \le i \le I$) is bounded by a polynomial in $(nd)$.
\item The degree of $\chi(\mathbf{y})$ is bounded by a polynomial in $(nd)$.
\item The height of $\chi(\mathbf{y})$, i.e. the largest coefficient in the polynomials in $\chi$ is bounded by $2^{\size({\tau(\mathbf{y})})(nd)^{\mathcal{O}(1)}}$
\end{enumerate}
Moreover, the procedure runs in time polynomial in the size of the input formula $\tau(\mathbf{y})$.
\end{thm}

\begin{prop}
\label{prop:Renegar}
$\mu,\nu$ are algebraic and computable (in finite time). Further, if $K$ is a bound on the order of the LRS: (a) this computation runs in polynomial time, and furthermore, (b) if $\mu, \nu$ are nonzero, we have $|\mu|,|\nu| < 2^{s^{\mathcal{O}(1)}}$ and $\frac{1}{|\mu|},\frac{1}{|\nu|} < 2^{s^{\mathcal{O}(1)}}$
\end{prop}

\begin{proof}
  We obtain the statement directly from Renegar \cite{Renegar}, i.e. $\mu$, $\nu$ are roots of polynomials obtained via quantifier elimination on the following formulae that can be expressed in the First Order Theory of Reals.  $\mu$ is the unique satisfying assignment to
 \begin{equation}
 \label{eq:fornu}
\left(\forall \vect t.~ \vect t \in T \Rightarrow \dominant(\vect c_0, \vect t) \ge \mu\right) \land \left(\exists \vect t_0 \in T.~ \dominant(\vect c_0, \vect t_0) = \mu\right)
 \end{equation}
 and is hence algebraic, with degree and height constrained by the formula.
 
Note that $|x| \ge b$ can be expressed as $(x \ge b \lor x \le -b)$. Similarly, $|x| = b$ can be expressed as $b \ge 0 \land (x = b \lor x = -b)$. $\nu$ is the unique satisfying assignment to
\begin{equation}
\label{eq:formu}
\left(\forall \vect t.~ \vect t \in T \Rightarrow |\dominant(\vect c_0, \vect t)| \ge \nu\right) \land \left(\exists \vect t_0 \in T.~ |\dominant(\vect c_0, \vect t_0)| = \nu\right).
 \end{equation}
 
Computability follows by performing quantifier elimination on the above formulae, which, as discussed in the proof of Proposition \ref{prop.ultimate}, will have size polynomial in that of the input. This yields polynomial equations, of which $\mu$ (resp.\ $\nu$) are roots, and hence algebraic. 

Now, if the order of the LRS is bounded by $K$, then this ensures that formulae (\ref{eq:fornu}) and (\ref{eq:formu}) use boundedly many variables (depending on $K$). This is seen by generalising the arguments made in \cite[Section 3.1]{ouaknine2014positivity}. Essentially, from the bound on order one can derive bounds on the dimension of the torus T and the number of constraints that define it, and thus the number of variables. Hence, we can apply Theorem \ref{theorem:renegar}. From this the complexity bounds on $\mu, \nu$ follow as quantifier elimination runs in polynomial time under these assumptions. The equivalent quantifier free formula, and hence $\mu, \nu$ themselves have degree polynomial and height single exponential in the size of the original formula (and thus the input).
\end{proof}
}

For $\exists$-robust Skolem we adapt the above argument but with $\nu$ instead of $\mu$. We start with the following useful property about $\nu$.
 
\begin{prop}
   \label{prop:nuprop}
 If $\nu = 0$, then $\forall \varepsilon>0$,
$\exists \vect c_\varepsilon, n$ with 
$|\vect c_0 - \vect c_\varepsilon|\leq \varepsilon$ such that $v_n(\mathbf{c}_{\varepsilon}) = 0$.
\end{prop}

\begin{proof}
  Assume $\nu = 0$, and let $\varepsilon>0$.
We again prove that 
for all $\alpha>0$, we have a $n_\alpha$ such that 
$|v_{n_\alpha}(\vect c_0)| \leq \alpha$, which suffices by Lemma
\ref{l.distance}.
Let $\alpha>0$ arbitrarily small, and let $N$ such that for all $n>N$, 
$|v_n^{res}(\vect c_0)| \leq \frac{\alpha}{2}$. This $N$ exists
as $v_n^{res}(\vect c_0) \rightarrow_{n \rightarrow \infty} 0$.
By Kronecker, as $\nu=0$, there exists $n_{\alpha}>N$ with 
$|v^{dom}_{n_{\alpha}}(\vect c_0)| < \frac{\alpha}{2}$.
Thus $|v_{n_{\alpha}}(\vect c_0)| \leq |v_{n_{\alpha}}^{dom}(\vect c_0)| + |v_{n_{\alpha}}^{res}(\vect c_0)| \leq \alpha$. 
\end{proof}

\begin{algorithm}[t]
\caption{to check $\exists$-robust Skolem of initial configuration $\vect c_0$}
\label{algoS}
\SetAlgoLined
\DontPrintSemicolon
\KwData{Companion matrix $\M \in \mathbb{Q}^{\kappa \times \kappa}$
of $\lrs{u}$ and initial configuration $\vect c_0 \in \mathbb{Q}^{\kappa}$}

$\{\gamma_j\}_j \gets$ eigenvalues of $\M$, \quad $\rho \gets \max_j |\gamma_j|$, \quad
$\{e^{i \theta_j}\}_{j=1}^k \gets \{\gamma_i/\rho ~|~ |\gamma_i| = \rho\}$ \;

Determine $T$ Torus obtained by applying Masser's result (Theorem \ref{thm:abelian}) to $\{\theta_j\}_{j=1}^k$ \;

$\nu \gets \min_{\vect t \in T}|\dominant(\vect c_0, \vect t)|$ (Proposition \ref{prop:Renegar}) \;

\eIf{$\nu = 0$}
{\Return NO (Proposition \ref{prop:nuprop})}
{
Compute $N_{thrSk}$ such that $|v_n^{res}(\vect c_0)| < \frac{\nu}{2}$ for all $n> N_{thrSk}$\;

\ForEach{$n \in \{0, 1, \dots, N_{thrSk}\}$}
{
\If{$\mathbf{M}^n \vect c_0 = 0$}
{
\Return NO \;
}
}
\Return YES \;
}
\end{algorithm}

The procedure to decide $\exists$-robust Skolem is given in Algorithm~\ref{algoS}. Basically, we compute $\nu \gets \min_{\vect t \in T}|\dominant(\vect c, \vect t)|$ using Proposition \ref{prop:Renegar}, for $\vect c_0$ the initial configuration around which we are looking for a neighbourhood. If $\nu=0$, we declare $\exists$-robust Skolem does not hold,
according to Proposition \ref{prop:nuprop}.

Otherwise, we compute $N_{thrSk}$ such that $|v_n^{res}(\vect c_0)| < \frac{\nu}{2}$ for all $n> N_{thrSk}$. Then we check if $v_{n}(\vect c_0) = 0$ for some $n \leq N_{thrSk}$. If yes, then $\exists$-robust Skolem does not hold. Otherwise we know that 
$\exists$-robust Skolem holds. 

Indeed, if $\nu>0$, 
for all $n>N_{thrSk}$, $|v_n(\vect c_0)| = |v^{dom}_n(\vect c_0) + v^{res}_n(\vect c_0)
|> \nu - \frac{\nu}{2} \geq \frac{\nu}{2} >0$, and this remains $>0$ in a neighbourhood of $\vect c_0$. 
Similarly, by hypothesis, $|v_n(c_0)|>0$ for the finite number of $n< N_{thrSk}$,
and in particular we have a lower bound $>0$ that ensure it stays strictly positive in a neighbourhood of $c_0$.

The complexity follows since both $\nu$ and $1/\nu = 2^{s^{\Oone}}$ are bounded (Proposition \ref{prop:Renegar}) and thus have $N_{thrSk} = 2^{s^{\Oone}}$ since $v^{res}_n(\vect c_0) = \OO(\frac{1}{n})$. %

This finally completes the proof of Theorem \ref{th.robust}.

\section{Conclusion}
\label{sec:conclusion}

We have formulated a natural notion of robustness for the Skolem and (ultimate) positivity problems and shown several results: for a given neighbourhood around an initial configuration $\vect c_0$, we show Diophantine hardness for the problems. Interestingly, this is the first Diophantine hardness result for a variant of Skolem as far as we know. This implies that for uninitialised positivity, the fact that the initial configuration $\vect c_0$ is arbitrary is crucial to obtain decidability \cite{Tiwari04,Braverman06}, as having a fixed ball around $\vect c_0$ is not sufficient. This is the case for all considered problems except the non-uniform variant of ultimate positivity for open balls, where we obtain decidability. Our results are for rational LRRs with the initial configuration having real-algebraic entries. We leave open hardness for the case of rational LRRs where the initial configuration has only rational entries.

We proved decidability of $\exists$-robust Skolem and (ultimate) positivity problems around an initial configuration in full generality {with real-algebraic entries}. These questions are also arguably more practical as in a real system, it is often impossible to determine the initial configuration with absolute accuracy. Our results can provide a precision with which it is sufficient to set the initial configuration. Beyond these technical results, we provided geometrical reinterpretations of Skolem/positivity, shedding a new light on this hard open problem.

\bibliography{biblio}

\appendix

\section{}

\label{sec:app-number-theory}

In this Appendix, we make a few remarks regarding the representation of numbers. A complex number $\alpha$ is said to be algebraic if it is a root of a polynomial with integer coefficients. For an algebraic number $\alpha$, its defining polynomial $p_\alpha$ is the unique polynomial of least degree of $\zx$ such that the GCD of its coefficients is $1$ and $\alpha$ is one of its roots.
Given a polynomial $p\in \zx$, we denote the length of its representation $\size(p)$, its height $H(p)$ the maximum absolute value of the coefficients of $p$ and $d(p)$ the degree of $p$. When the context is clear, we will only use $H$ and $d$.

A separation bound provided in \cite{mignottecon} has established that for distinct roots $\alpha$ and $\beta$ of a polynomial $p \in \mathbb{Z}[X]$,
$|\alpha - \beta| > \frac{\sqrt{6}}{d^{(d+1)/2}H^{d-1}}$.
This bound allows one to represent an algebraic number $\alpha$ as a 4-tuple $(p,a,b,r) \in \mathbb{Z}[X]\times \mathbb{Q}^3$ where $\alpha$ is the only root of $p$ at distance $< r$ from $a+ib$, and we denote $\size({\alpha})$ the size of this representation, i.e., number of bits needed to write down this 4-tuple.

Further, we note that two distinct algebraic numbers $\alpha$ and $\beta$, are always roots of $p_\alpha p_\beta$, and we have that 
\begin{equation}
\frac{1}{|\alpha - \beta|} = 2^{(||\alpha||+||\beta||)^{\Oone}}
\label{eq:mignottebound}
\end{equation}

Given a polynomial $p\in \zx$, one can compute its roots in polynomial time wrt $\size(p)$ \cite{findroots1operate1}. Since algebraic numbers form a field, given $\alpha$, $\beta$ two algebraic numbers, one can always compute the representations of $\alpha+\beta$, $\alpha\beta$, $\frac 1 \alpha$, $Re(\alpha)$, $Im(\alpha)$ in polynomial time wrt $\size(\alpha)+\size(\beta)$ \cite{findroots1operate1,findroots2operate2}.

\end{document}